\documentclass[%
 reprint,
showpacs,preprintnumbers,
 amsmath,amssymb,
 aps,
prc,
superscriptaddress]{revtex4-1}

\usepackage{graphicx}
\usepackage{dcolumn}
\usepackage{bm}

\begin{document}

\preprint{APS/124-QCD}

\title{Effects of the Symmetry Energy and its Slope on Neutron Star Properties}

\author{Luiz L. Lopes}
\email{llopes@varginha.cefetmg.br}

\affiliation{%
 Departamento de Fisica, CFM - Universidade Federal de Santa Catarina;  C.P. 476, CEP 88.040-900, Florian\'opolis, SC, Brasil 
}%
\affiliation{%
 Centro Federal de Educa\c{c}\~ao Tecnol\'ogica de Minas Gerais Campus VIII; CEP 37.022-56, Varginha - MG - Brasil
}%

\author{Debora P. Menezes}%

\affiliation{%
 Departamento de Fisica, CFM - Universidade Federal de Santa Catarina;  C.P. 476, CEP 88.040-900, Florian\'opolis, SC, Brasil 
}%

\date{\today}

\begin{abstract}
In this work we study the influence of the symmetry energy and its slope on three major properties of neutron stars: the maximum mass, the radii
of the canonical 1.4$M_\odot$  and the minimum mass that enables the direct URCA effect.
We utilize four parametrizations  of the relativistic quantum hadrodynamics and  vary 
the symmetry energy within accepted  values. We see that although the maximum mass is almost independent of it, the radius of the canonical $1.4M_\odot$
and the mass that enables  the direct URCA effect is strongly correlated with the symmetry energy and its slope. Also, since we expect that the radius grows with
the slope, a theoretical limit  arises when we increase  this quantity above certain values.
\end{abstract}


\maketitle

\section{Introduction}

Although   nuclear matter properties are well known around the  saturation point, the physics of very high density and strongly isospin-asymmetric matter is
far from being completely understood. This extreme region is important to determine  the main properties of an exotic object, the neutron star.  
Neutron stars are compact objects maintained by the equilibrium of gravity and the degenerescence pressure of the fermions together with a strong nuclear repulsion force due to the high density reached  in their interior.  

In the present work we focus on the properties of nuclear matter at sub-threshold density,
 which we describe with the relativistic quantum hadrodynamics (QHD)~\cite{Serot}.
 QHD is an effective model where the strong interaction is simulated by the exchange of massive mesons through Yukawa potentials.
 In the present work, we use the scalar-isoscalar $\sigma$ meson, and the vector-isoscalar $\omega$ meson to describe the properties of symmetric nuclear matter,
and  the vector-isovector $\rho$ meson  to correct the value of the  symmetry energy~\cite{Glen0} and
 describe effects of isospin-asymmetric matter. This $\sigma\omega\rho$ model is the standard model of QHD in the current literature~\cite{Glen,schmitt,Max}.
 Within this model, once  the coupling constant of the $\rho$ meson is fixed the symmetry energy and its slope are established.
 Since the slope of the symmetry energy is important to constrain the neutron stars radii~\cite{Rafa2011},  another parameter is necessary if one wants 
to vary the slope without varying the symmetry energy and vice-versa. To accomplish this task,
 we include the scalar-isovector $\delta$ $[a_0(980)]$ meson in a more
complete $\sigma\omega\rho\delta$ model~\cite{Kubis,Liu}. The effects of the $\delta$ meson in asymmetric matter were already
 studied in several topics as the neutron radii~\cite{Fur}, linear response~\cite{Kubis,Liu,Greco}
 and even neutron star properties~\cite{Debdelta,Liu2}, however, in a different approach.
 Instead of worrying about the strength of the coupling constant of the delta meson with the baryons~\cite{Liu,Debdelta,Liu2},
 what is strongly model dependent,  we focus on the fitting of the physical quantities given by the  symmetry energy and its slope.

{ There are many different parametrizations for the QHD models in the literature, all of them chosen so as to reproduce nuclear matter bulk properties. In a recent work \cite{Dutra}, an extensive review of 
$263$ parametrizations of different types of RMF models were analysed under three different 
sets of constraints related to symmetric nuclear matter, pure neutron matter, 
symmetry energy, and its derivatives. In this paper, we utilize four of these parametrizations: GM1, 
GM3~\cite{Glen2}, NL$\rho$~\cite{Liu} and NL3~\cite{Lala} to describe the properties of 
symmetric nuclear matter. Not all of them satisfy the constraints investigated in \cite{Dutra}, but we perform
some modifications on the usual parameters based on the symmetry energy and its slope and in the conclusions section, we discuss possible constraints related to neutron star observational properties. }
The value of the symmetry energy at saturation density is well known to lie  between 30 MeV and 35 MeV~\cite{Tsang,Sun,Carbone}.
 The value of the symmetry energy slope is rather more controversial. Although some results point towards a very low value of slope, 
lower than 62 MeV~\cite{Lim,Steiner}, other studies indicate a much higher limit, up to 113 MeV~\cite{Tsang,Chen}. We follow this last prescription which is
very close to the limit of 115 MeV presented in Ref.~\cite{Dutra}.
Nevertheless, we can find in the literature values as high as 150 MeV~\cite{Debdelta} or even higher than 170 MeV, as pointed out
in a recent work~\cite{Cozma}. 

We study the influence of the energy symmetry and its slope on three major properties of neutron stars: the maximum mass, the radius
of the canonical $1.4M_\odot$ and the minimum mass that enables direct URCA effect. To  investigate these effects we utilize three different approaches.
First,  within the traditional $\sigma\omega\rho$ models, we vary the symmetry between acceptable values.  After, in order to investigate the individual 
effects of symmetry energy and the slope, we fix $L$ and vary  $S_0$ within the $\sigma\omega\rho\delta$ models.
 And then we  perform the inverse  situation, fixing $S_0$ and varying $L$.

This paper is organized as follows: we review  the formalism of the QHD models with and without the $\delta$ meson and the parametrizations
used in this work.
Then we present the numerical results for the three approaches and discuss the implications and  validity of the results. 
Finally we present the conclusions of our work.

\section{The Formalism}

We use an extended version of the relativistic QHD~\cite{Serot}, whose Lagrangian density reads:
\begin{widetext}
\begin{eqnarray}
\mathcal{L}_{QHD} =  \bar{\psi}_N[\gamma^\mu(i\partial_\mu  - g_{v}\omega_\mu  - g_{\rho} \frac{1}{2}\vec{\tau} \cdot \vec{\rho}_\mu)
- (m_N - g_{s}\sigma - g_{\delta}\vec{\tau} \cdot \vec{\delta})]\psi_N      
  + \frac{1}{2}(\partial_\mu \sigma \partial^\mu \sigma - m_s^2\sigma^2) \nonumber \\ + \frac{1}{2}(\partial_\mu \vec{\delta} \partial^\mu \vec{\delta} - m_{\delta}^2\delta^2)   - U(\sigma)   
   + \frac{1}{2} m_v^2 \omega_\mu \omega^\mu  + \frac{1}{2} m_\rho^2 \vec{\rho}_\mu \cdot \vec{\rho}^{ \; \mu}  - \frac{1}{4}\Omega^{\mu \nu}\Omega_{\mu \nu} -  \frac{1}{4}\bf{P}^{\mu \nu} \cdot \bf{P}_{\mu \nu}  , \label{s1} 
\end{eqnarray}
\end{widetext}
where  $\psi_N$  are the  baryonic  Dirac fields of the nucleons, and $\sigma$, $\omega_\mu$, $\vec{\delta}$ and $\vec{\rho}_\mu$ are the mesonic fields.
 The $g's$ are the Yukawa coupling constants that simulate the strong interaction, $m_N$ is the mass of the nucleon  (that we assume next as 939 MeV), $m_s$, $m_v$, $m_\delta$ and $m_\rho$ are
 the masses of the $\sigma$, $\omega$, $\delta$ and $\rho$ mesons respectively.
 The antisymmetric mesonic field strength tensors are given by their usual expressions as presented in~\cite{Glen}.
  The $U(\sigma)$ is the self-interaction term introduced in ref.~\cite{Boguta} to fix some of the saturation properties of the nuclear matter and is given by:

\begin{equation}
U(\sigma) =  \frac{1}{3!}\kappa \sigma^3 + \frac{1}{4!}\lambda \sigma^{4} \label{s2} .
\end{equation}
 
 Finally, $\vec{\tau}$ are the Pauli matrices. In order to describe a neutral, beta stable nuclear matter, we add leptons as free Fermi gases:
 
 \begin{equation}
 \mathcal{L}_{lep} = \sum_l \bar{\psi}_l [i\gamma^\mu\partial_\mu -m_l]\psi_l , \label{s3}
 \end{equation}
The electron and muon masses are 0.511 MeV and 105.6 MeV respectively.

To solve the equations of motion, we use the mean field approximation (MFA), where the meson fields are replaced by their expectation values, i.e:
  $\sigma$ $\to$ $\left < \sigma \right >$ = $\sigma_0$, $\delta$ $\to$ $\left < \delta \right >$ = $\delta_0$,  $\omega^\mu$ $\to$ $\delta_{0 \mu}\left <\omega^\mu  \right >$ = $\omega_{0}$  and
   $\rho^\mu$ $\to$ $\delta_{0 \mu}\left <\rho^\mu  \right >$ = $\rho_{0}$.
 The MFA gives us the following eigenvalue for the nucleon energy~\cite{Glen}:

\begin{equation}
E_N = \sqrt{k^2 + M^{*2}_N} + g_{v}\omega_0 + g_{\rho} \frac{\tau_3}{2}  \rho_0 ,  \label{s4}
\end{equation}
where $M_N^*$ is the nucleon effective mass: 
\begin{equation}
M^{*}_N \dot{=}   m_N - g_{s}\sigma_0 - g_{\delta}\tau_3 \delta_0.
\end{equation}

We see that while the vector-isovector $\rho$ meson splits the energies, the scalar-isovector $\delta$ meson splits the masses of the nucleons.
The third Pauli matrix $\tau_3$ assumes
the value +1 (-1) for  protons (neutrons).

For the leptons, the energy eigenvalues are those of the free Fermi gas:
\begin{equation}
\quad E_l = \sqrt{k^2 + m^{2}_l} , \label{s5}
\end{equation}
and the meson fields become:

 \begin{equation}
\omega_0  =  \frac{g_{v}}{m_v^2} (n_p + n_n) , \label{s6}
\end{equation}

\begin{equation}
\delta_0 =  \frac{g_{\delta}}{m_\delta^2}  (n_{Sp} - n_{Sn}), \label{s7} 
\end{equation}

\begin{equation}
\rho_0  =  \frac{g_{\rho}}{m_\rho^2} \frac{1}{2} (n_p - n_n),  \label{s8}
\end{equation}

\begin{equation}
\sigma_0 =  \frac{g_{\sigma}}{m_s^2}  (n_{Sp} + n_{Sn}) - \frac{1}{2}\frac{\kappa}{m_s^2}\sigma_0^2 -\frac{1}{6}\frac{\lambda}{m_s^2} \sigma_0^3 , \label{s9}
\end{equation}
where $n_{Sp}$, $n_{Sn}$ are the scalar densities,  and $n_{p}$ and $n_{n}$ are the number  densities of the protons and the neutrons respectively { and are given by}: 

\begin{eqnarray}
\quad n_{SB} =  \int_0^{k_{fB}} \frac{M^*_B}{\sqrt{k^2 + M^{*2}_B}} \frac{k^2}{\pi^2} dk,    \nonumber
 \\ n_B = \frac{k_{fB}^3}{3\pi^2}, \quad \mbox{and}  \quad  n= \sum_B n_B, \nonumber \\ \quad B = (p,n).
\label{s10}
\end{eqnarray}

To describe the properties of the nuclear matter, we calculate the EoS from statistical mechanics~\cite{Huang}. The nucleons and leptons,
 being fermions, obey the Fermi-Dirac distribution. In order to compare our results with experimental and 
observational constraints, we next study nuclear and stellar systems at zero temperature.
In this case the Fermi-Dirac distribution becomes the Heaviside step function. The  energy densities of  baryons, leptons and  mesons  read:

\begin{equation}
\epsilon_B =  \frac{1}{\pi^2} \sum_B \int_0^{k_f} \sqrt{k^2 + M^{*2}_B} k^2 dk , \label{s11}
\end{equation}
\begin{equation}
\epsilon_l = \frac{1}{\pi^2} \sum_l \int_0^{k_f} \sqrt{k^2 + m^{2}_l} k^2 dk , \label{s12}
\end{equation}
\begin{equation}
\epsilon_m =  \frac{1}{2}\bigg ( m_s^2\sigma_0^2 + m_v^2\omega_0^2 +m_\delta^2\delta_0^2 + m_\rho^2\rho_0^2 \bigg ) + U(\sigma) , \label{s13}
\end{equation}
where $k_f$ is the Fermi momentum, and we have already used the fact that the fermions have degeneracy  equal to 2. The total energy density is the sum of the partial ones:

\begin{equation}
\epsilon =   \epsilon_B +   \epsilon_l +  \epsilon_m ,\label{s14}
\end{equation}
and the pressure is calculated via thermodynamic relations:

\begin{equation}
P = \sum_f \mu_f n_f - \epsilon , \label{s15}
\end{equation}
where the sum runs over all the fermions ($f = B,l$) and $\mu$ is the chemical potential, which
corresponds exactly to the energy eigenvalue at $T=0$.

Now we couple the equations  imposing $\beta$  equilibrium and zero total net charge:

\begin{equation}
\mu_{p} = \mu_n -  \mu_e, \quad  \mu_e = \mu_\mu,   \quad  n_p + \sum_l  n_l =0 \label{s16} ,
\end{equation}
where $\mu_{p}$, $\mu_n$, $\mu_e$ and $\mu_\mu$ 
are the chemical potentials of the proton, neutron, electron and muon respectively.

The equation of states developed in this work are used as input to solve the Tolman-Oppenheimer-Volkoff (TOV) equations~\cite{TOV},
that describe  a static, spherically symmetric, relativistic star in hydrostatic equilibrium.
 The neutron star crust is simulated by the BPS EoS~\cite{BPS}.

To compare the different approaches we need the symmetry energy of the symmetric nuclear
matter ($n_p=n_n$), which  is defined  as~\cite{Rafa2011,Kubis,Liu}:

\begin{eqnarray}
S = \frac{1}{8} \bigg ( \frac{g_\rho}{m_\rho} \bigg )^2 + \frac{k_f^2}{6 \sqrt{k_f^2 + M^{*2}}} \nonumber \\
 - \bigg (\frac{g_\delta}{m_\delta} \bigg )^2
\frac{M^{*2}n}{2(k_f^2 + M^{*2})[1 + (g_\delta/m_\delta)^2 \cdot A(k_f)]}    , \label{s17} 
\end{eqnarray}
where

\begin{equation}
 A(k_f) = \frac{4}{(2\pi)^3}\int_{0}^{k_f} d^3k \frac{k^2}{k^2 + M^{*2}} , \label {s18}
\end{equation}
is a function of the Fermi momentum ($k_f = k_{fp} = k_{fn}$) and the effective mass 
($M^{*} = M^{*}_p = M^{*}_n$)
of symmetric nuclear matter. According to Eq.(\ref{s17}), to fit the bulk properties of nuclear matter,
we are not able to fix  $(g_\rho/m_\rho)^2$ and $(g_\delta/m_\delta)^2$ independently. We then explore
a family of values that allow the symmetry energy to lie between 30 and 35 MeV, { bearing in mind that a}
 maximum value  $(g_\delta/m_\delta)^2$ = 2.6 $fm^{-2}$
arises from the so called $BonnC$ potential~\cite{Kubis,Liu,Machleidt}.

It is useful to expand the symmetry energy $S$ around the saturation density ($n_0$) in a Taylor series as~\cite{Tsang}:
 
\begin{equation}
 S = S_0 + L\epsilon + O(\epsilon)^2 , \label{s19}
\end{equation}
where $S_0$ is the symmetry energy at nuclear saturation point, and  $\epsilon = (n_0 -n)/3n_0$.
 The parameter $L$ is the so called slope of the symmetry energy, and is calculated at 
nuclear saturation density as:
\begin{equation}
 L =   3n\bigg ( \frac{dS}{dn} \bigg ) \bigg |_{n = n_0} . \label{s20} 
\end{equation}

We can also define the slope for an arbitrary density $L(n)$ as~\cite{X}:

\begin{equation}
L(n) =    3n\bigg ( \frac{dS}{dn} \bigg ). \label{s21}
\end{equation}

\section{Model Parameters and results}

We utilize four well-known QHD parametrizations to fit the properties of nuclear saturation density.
The nuclear saturation density, $n_0$ ; binding energy per baryon, $B/A$, the effective nucleon mass, $M^{*}$
and the nuclear compression modulus $K$ are fixed parameters and are presented in Table \ref{T1} alongside
the symmetry energy $S_0$ and its slope $L$.

\begin{table}[ht]
\begin{center}
\begin{tabular}{|c||c|c|c|c|}
\hline 
  & GM1~\cite{Glen2} &  GM3~\cite{Glen2} & NL3~\cite{Lala} & NL$\rho$~\cite{Liu}  \\
 \hline
 $(g_s/m_s)^2$ ($fm^{-2}$) & 11.79  & 9.927 & 15.746 & 10.330  \\
 \hline
  $(g_v/m_v)^2$ ($fm^{-2}$) & 7.149 & 4.820 & 10.516 & 5.421\\
 \hline
  $(g_\rho/m_\rho)^2$ ($fm^{-2}$) & 4.410 & 4.791 & 5.360 & 3.830\\
 \hline
$\kappa/M_N$ & 0.005894 & 0.017318 & 0.0041014 &  0.01387 \\
\hline
$\lambda$ &  -0.006426 & -0.014526 & -0.015921 & -0.0288 \\
\hline  
\hline
$n_0$ $(fm^{-3})$ & 0.153 & 0.153 & 0.148 & 0.160 \\
\hline
$M^{*}/M$ & 0.70 & 0.78 & 0.60  & 0.75\\
\hline
K $(MeV)$ & 300 & 240 & 272 & 240 \\
\hline
$B/A$  $(MeV)$ & -16.3 & -16.3 & -16.3 & -16.0 \\ 
\hline
$S_0$  $(MeV)$ & 32.49 & 32.49 & 37.40 & 30.49 \\ 
\hline
$L$  $(MeV)$ & 93.7 & 89.7 & 118.4 & 85.0 \\ 
\hline
\end{tabular} 
\caption{Parameters and physical quantities for the original GM1, GM3, NL3 and NL$\rho$ models.} 
\label{T1}
\end{center}
\end{table}

Nevertheless, we consider the symmetry energy  and its slope  as free parameter and
 divide this subject in three parts.

\subsection{No $\delta$ meson}

We first   study the influence of $S_0$ and $L$ without the $\delta$ meson, where 
the $\rho$ meson determines simultaneously the symmetry energy and its slope. The parameters utilized
in this approach  are presented in Tables \ref{T2}, \ref{T3}, \ref{T4} and \ref{T5}  for GM1, GM3, NL3 and  NL$\rho$ respectively.
It is worth noting  that the original NL3~\cite{Lala} has the values of 37.4 MeV  and 118.4 MeV for $S_0$ and $L$ respectively.
 Both values are in disagreement with the experimental constraints~\cite{Tsang,Chen,Dutra}.
However, we can fix this problem by redefining the coupling constant of the $\rho$ meson, requiring 
that the symmetry energy assumes reasonable values.

From Tables \ref{T2} to \ref{T5}, we see that without $\delta$ meson, the symmetry energy slope shows a perfect linear dependence with $S_0$. This effect is independent
of the parametrization:

\begin{equation}
 L = 3S_0 + C , \label{s24}
\end{equation}
where the constant $C$ is model dependent, but the angular coefficient is not.

\begin{table}[!htb]
\begin{center}
\begin{tabular}{|c|c||c|c|}
\hline 
$(g_\rho/m_\rho)^2$ ($fm^{-2}$)  &  $(g_\delta/m_\delta)^2$ ($fm^{-2}$) &  $S_0$ (MeV)  & L (MeV)   \\
 \hline
 3.880 & -  & 30.49 & 87.9   \\
 \hline
 4.145 & - & 31.49 & 90.9 \\
 \hline
 4.410 & - & 32.49 & 93.9  \\
\hline
 4.677 & - & 33.49 & 96.9  \\
\hline  
 4.936 & - & 34.49 & 99.9  \\
\hline
\end{tabular} 
\caption{ $S_0$ and $L$ values for GM1 parametrization without $\delta$ meson} 
\label{T2}
\end{center}
\end{table}

\begin{table}[!htb]
\begin{center}
\begin{tabular}{|c|c||c|c|}
\hline 
$(g_\rho/m_\rho)^2$ ($fm^{-2}$)  &  $(g_\delta/m_\delta)^2$ ($fm^{-2}$) &  $S_0$ (MeV)  & L (MeV)   \\
 \hline
 4.260 & -  & 30.49 & 83.7   \\
 \hline
 4.525 & - & 31.49 & 86.7 \\
 \hline
 4.791 & - & 32.49 & 89.7  \\
\hline
 5.055 & - & 33.49 & 92.7  \\
\hline  
 5.319 & - & 34.49 & 95.7  \\
\hline
\end{tabular} 
\caption{$S_0$ and $L$ values for GM3 parametrization without $\delta$ meson} 
\label{T3}
\end{center}
\end{table}

\begin{table}[!htb]
\begin{center}
\begin{tabular}{|c|c||c|c|}
\hline 
$(g_\rho/m_\rho)^2$ ($fm^{-2}$)  &  $(g_\delta/m_\delta)^2$ ($fm^{-2}$) &  $S_0$ (MeV)  & L (MeV)   \\
 \hline
 3.458 & -  & 30.40 & 97.4   \\
 \hline
 3.733 & - & 31.40 & 100.4 \\
 \hline
 4.006 & - & 32.40 & 103.4  \\
\hline
 4.280 & - & 33.40 & 106.4  \\
\hline  
 4.552 & - & 34.40 & 109.4  \\
\hline
\end{tabular} 
\caption{ $S_0$ and $L$ values for NL3 parametrization without $\delta$ meson} 
\label{T4}
\end{center}
\end{table}

\begin{table}[!htb]
\begin{center}
\begin{tabular}{|c|c||c|c|}
\hline 
$(g_\rho/m_\rho)^2$ ($fm^{-2}$)  &  $(g_\delta/m_\delta)^2$ ($fm^{-2}$) &  $S_0$ (MeV)  & L (MeV)   \\
 \hline
 3.830 & -  & 30.49 & 85.0   \\
 \hline
 4.082 & - & 31.49 & 88.0 \\
 \hline
 4.335 & - & 32.49 & 91.0  \\
\hline
 4.558 & - & 33.49 & 94.0  \\
\hline  
 4.842 & - & 34.49 & 97.0  \\
\hline
\end{tabular} 
\caption{ $S_0$ and $L$ values for NL$\rho$ parametrization without $\delta$ meson} 
\label{T5}
\end{center}
\end{table}

\begin{figure*}[hb]
\begin{tabular}{cc}
\includegraphics[width=5.6cm,height=6.2cm,angle=270]{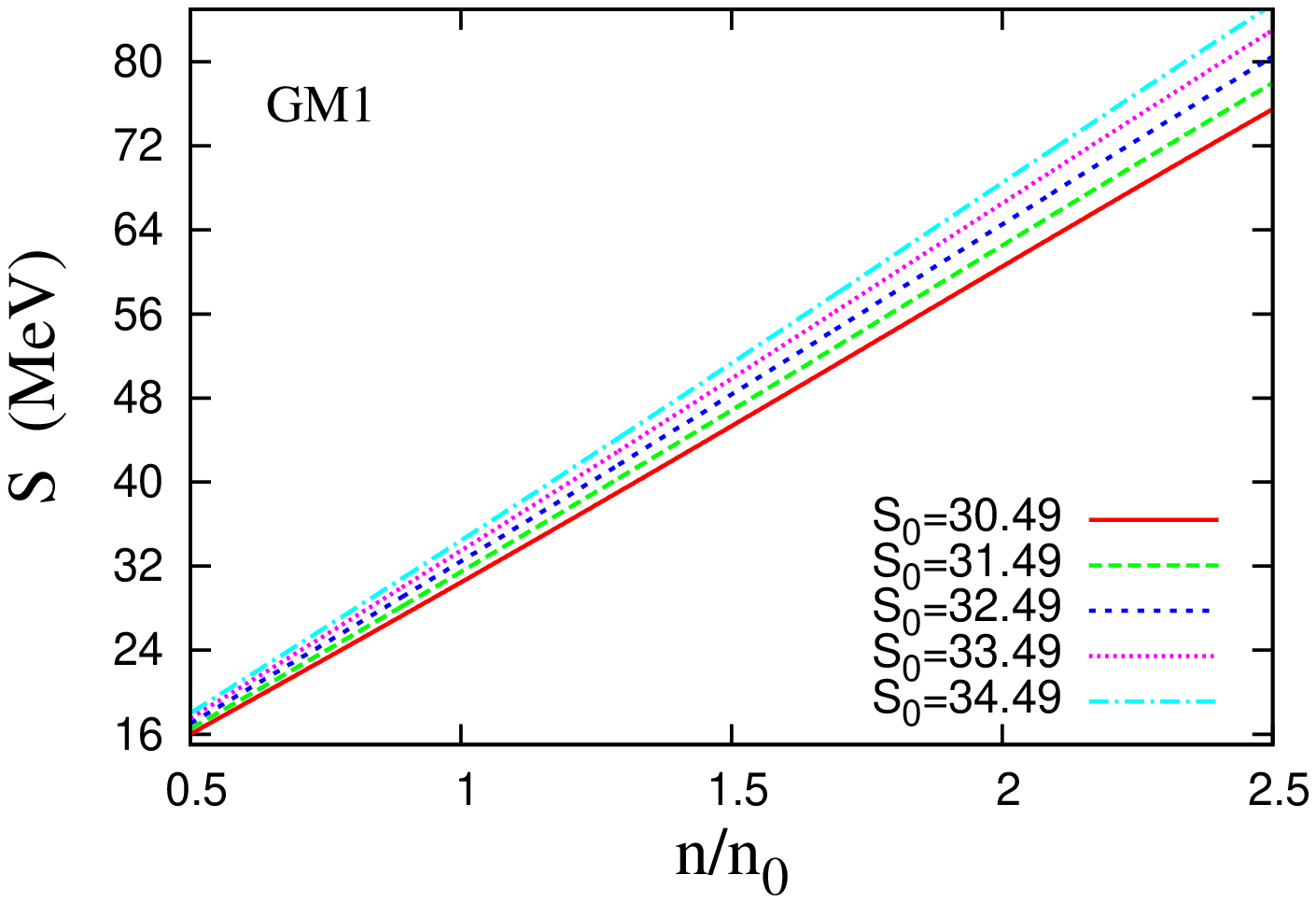} &
\includegraphics[width=5.6cm,height=6.2cm,angle=270]{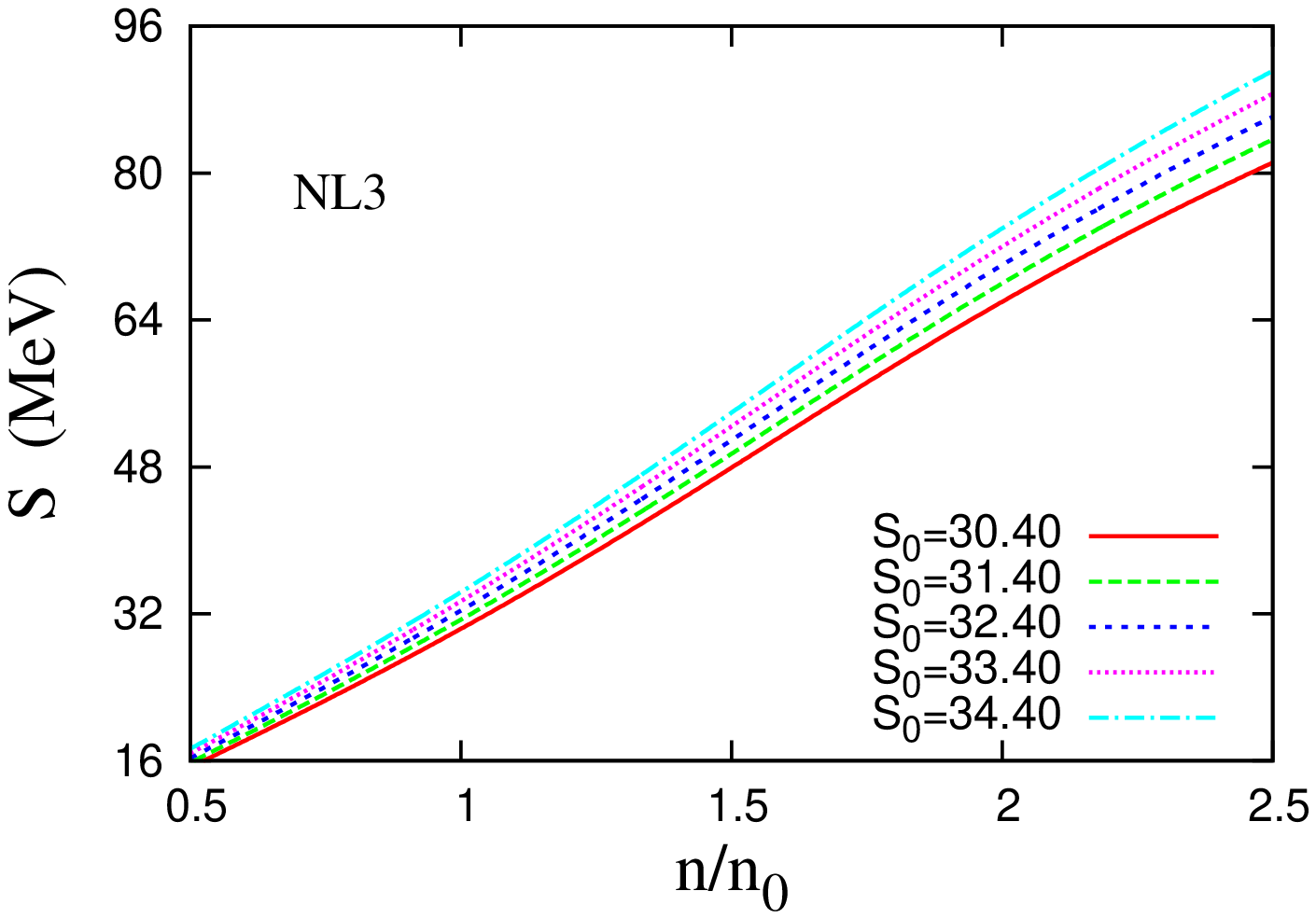} \\
\includegraphics[width=5.6cm,height=6.2cm,angle=270]{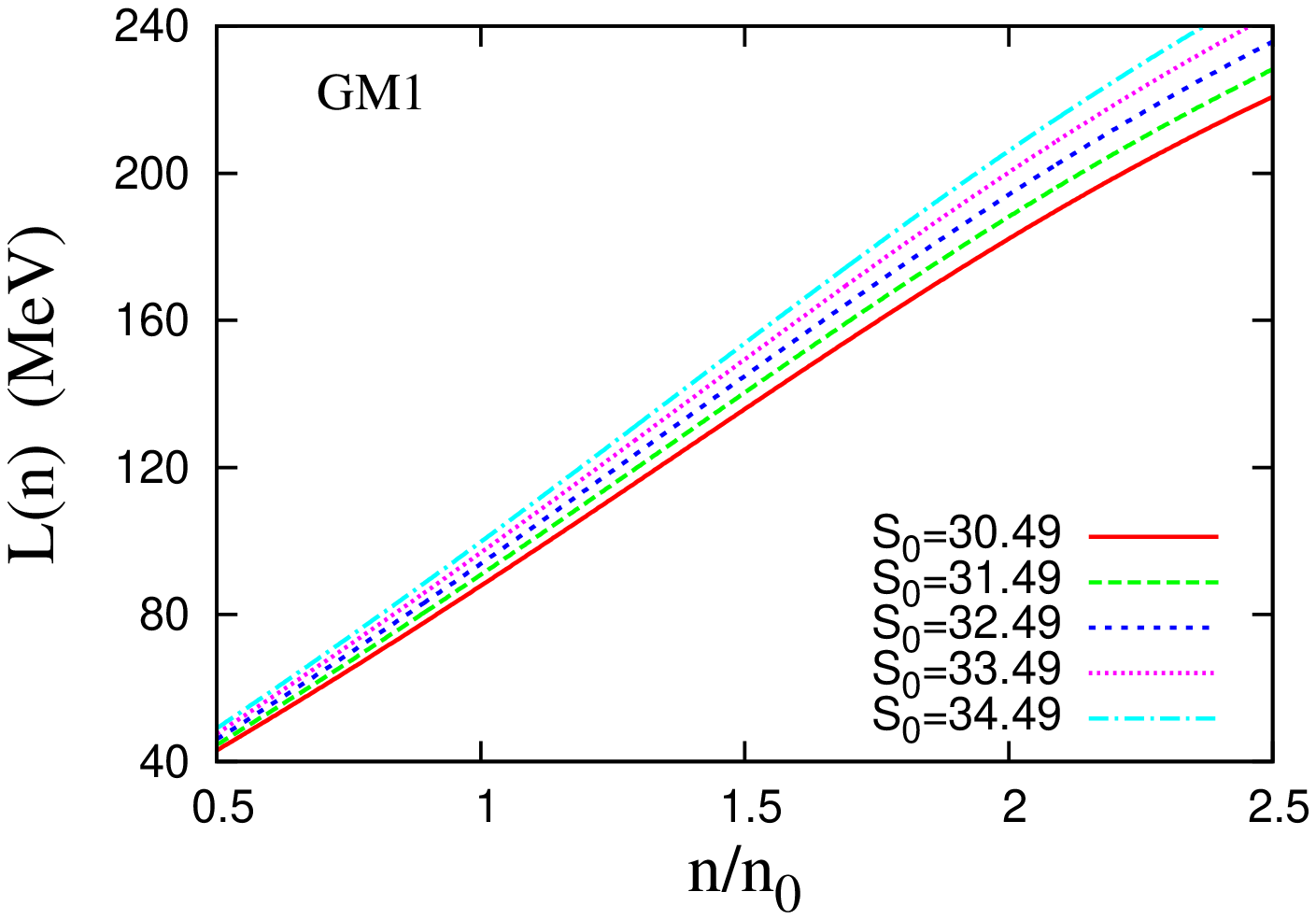} &
\includegraphics[width=5.6cm,height=6.2cm,angle=270]{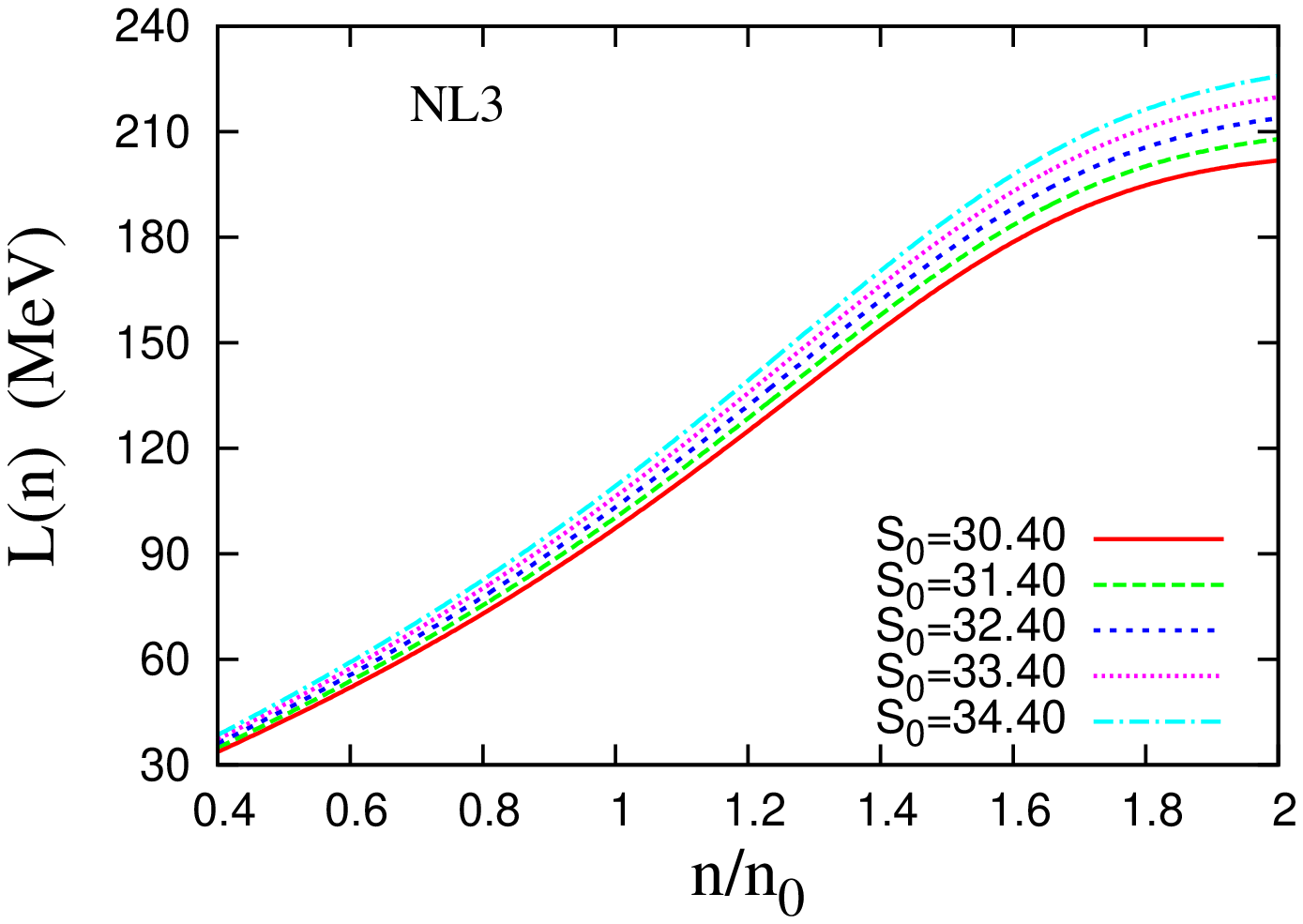} \\
\end{tabular}
\caption{(Color online) (Top) Symmetry energy $S$ as function of density and (Bottom) slope of the symmetry energy $L(n)$ as function of density with the $\sigma\omega\rho$ model. } \label{F1}
\end{figure*}

Without the $\delta$ meson, the variation of the symmetry energy $S_0$ implies a variation of the slope $L$.
The symmetry energy $S$ (Eq.(\ref{s17}))  and the slope for arbitrary densities
$L(n)$ (Eq.(\ref{s21})) for GM1 and NL3 are plotted in Fig.\ref{F1}.

We see that  the symmetry energy and the slope grows with the density and their lines never
cross each other. In other words, the parametrization with lower symmetry energy and slope at low densities  remains the parametrization with lower symmetry energy and slope at high densities. The behaviour of GM3 and NL$\rho$ are similar to those showed in Fig.~\ref{F1}.

\begin{figure*}[ht]
\begin{tabular}{cc}
\includegraphics[width=5.6cm,height=6.2cm,angle=270]{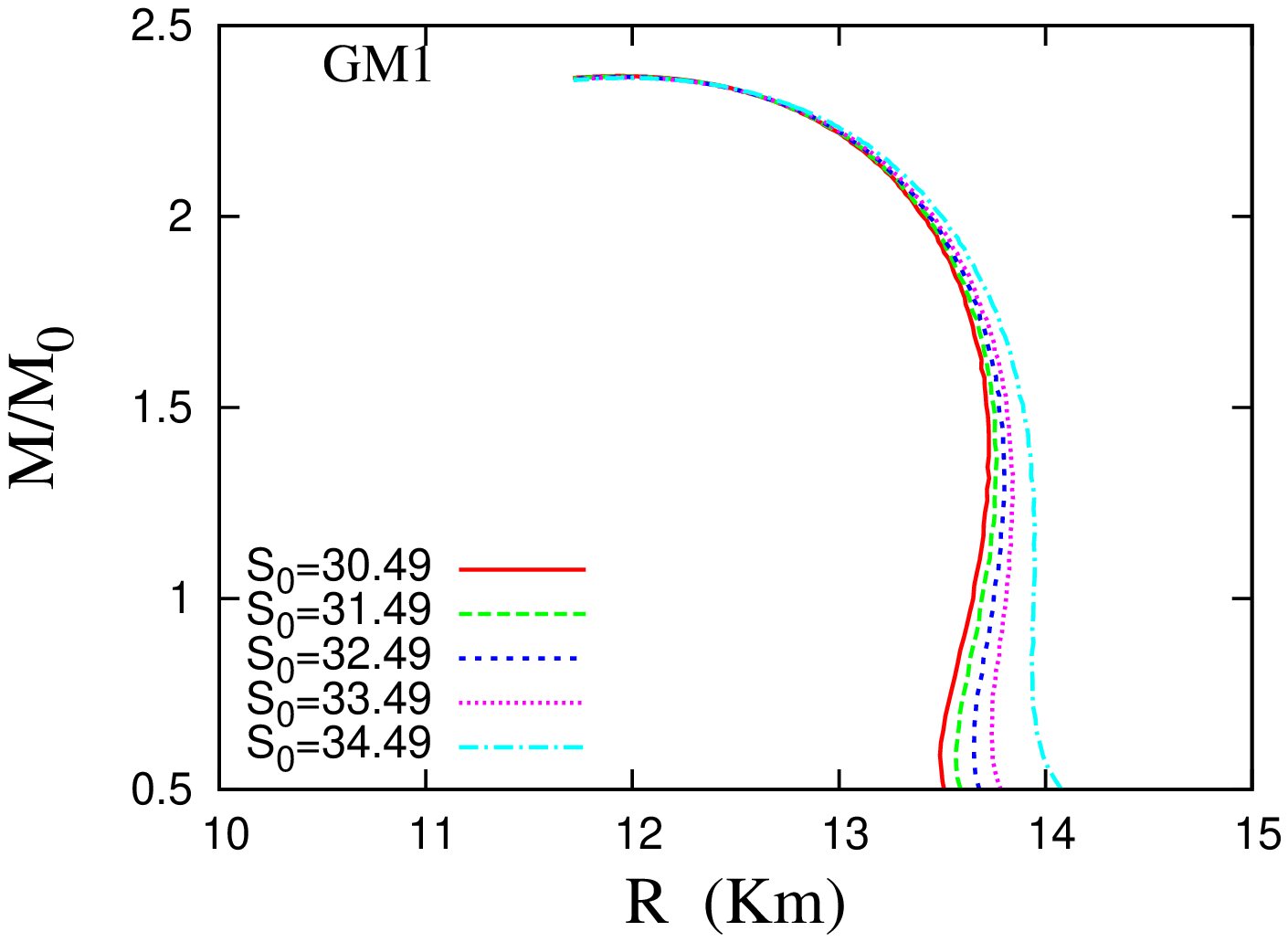} &
\includegraphics[width=5.6cm,height=6.2cm,angle=270]{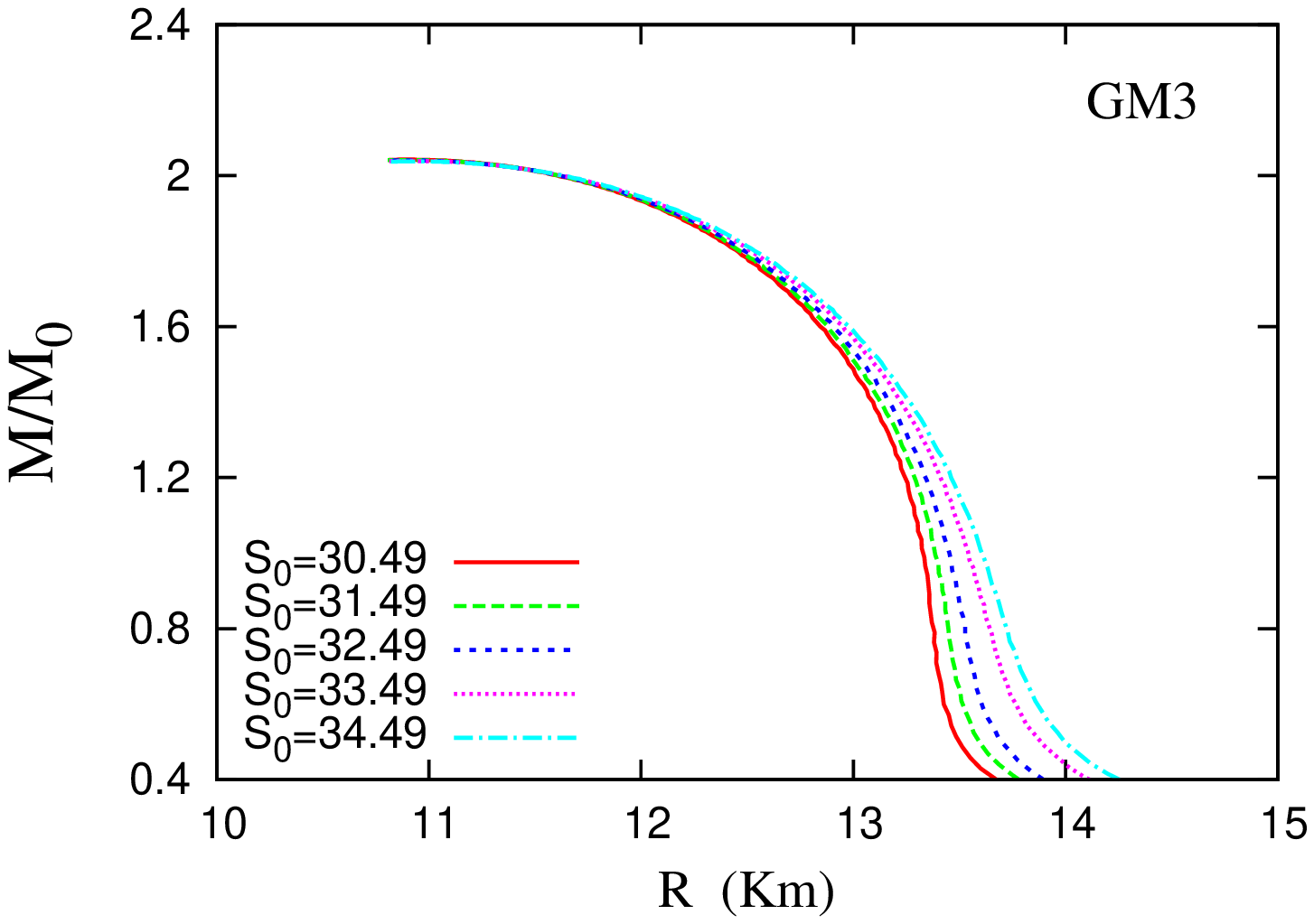} \\
\includegraphics[width=5.6cm,height=6.2cm,angle=270]{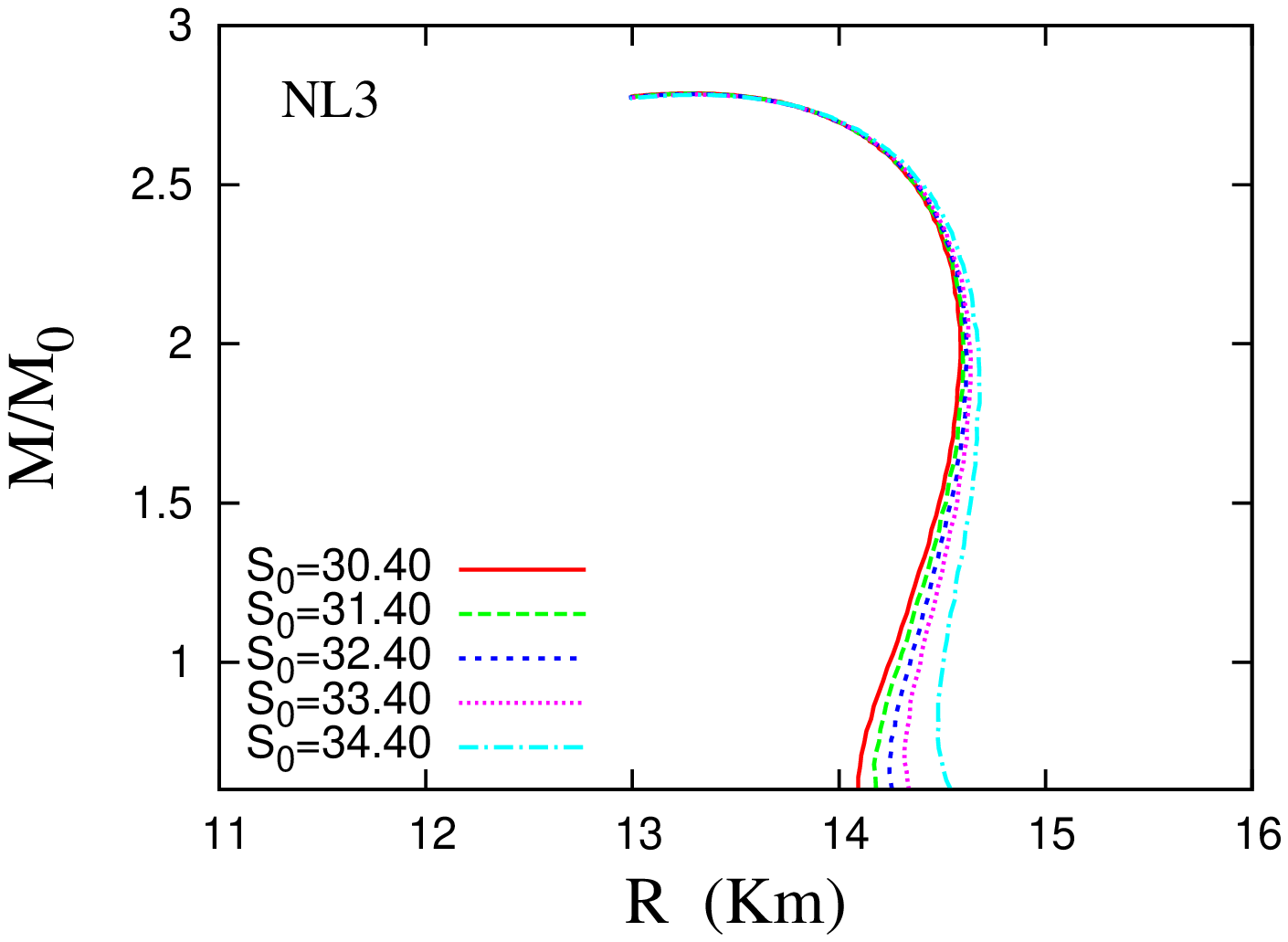} &
\includegraphics[width=5.6cm,height=6.2cm,angle=270]{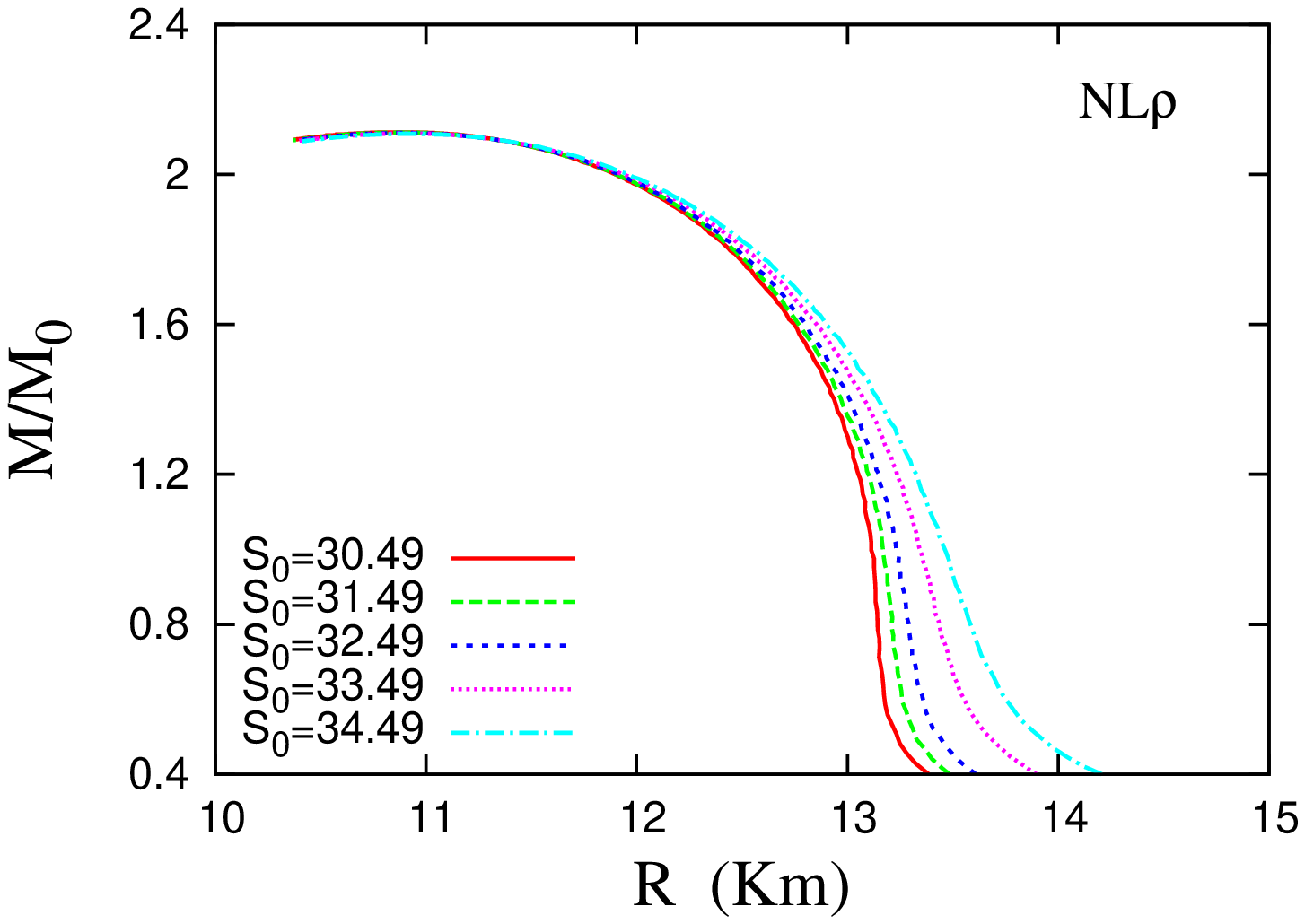} \\
\end{tabular}
\caption{(Color online) Neutron star mass-radius relation without $\delta$ meson. } \label{F3}
\end{figure*}

Now, we solve the TOV equations~\cite{TOV} and plot the results in Fig.\ref{F3}.
The symmetry energy and its slope have very little influence on the radii and
almost zero influence on the maximum masses. Indeed, the radii of the $1.4M_\odot$ varies about $0.2$ km
in all four parametrizations, always increasing with the symmetry energy and the slope,
 while the  the maximum masses vary  less than $0.02M_\odot$.

\begin{figure*}[ht]
\begin{tabular}{cc}
\includegraphics[width=5.6cm,height=6.2cm,angle=270]{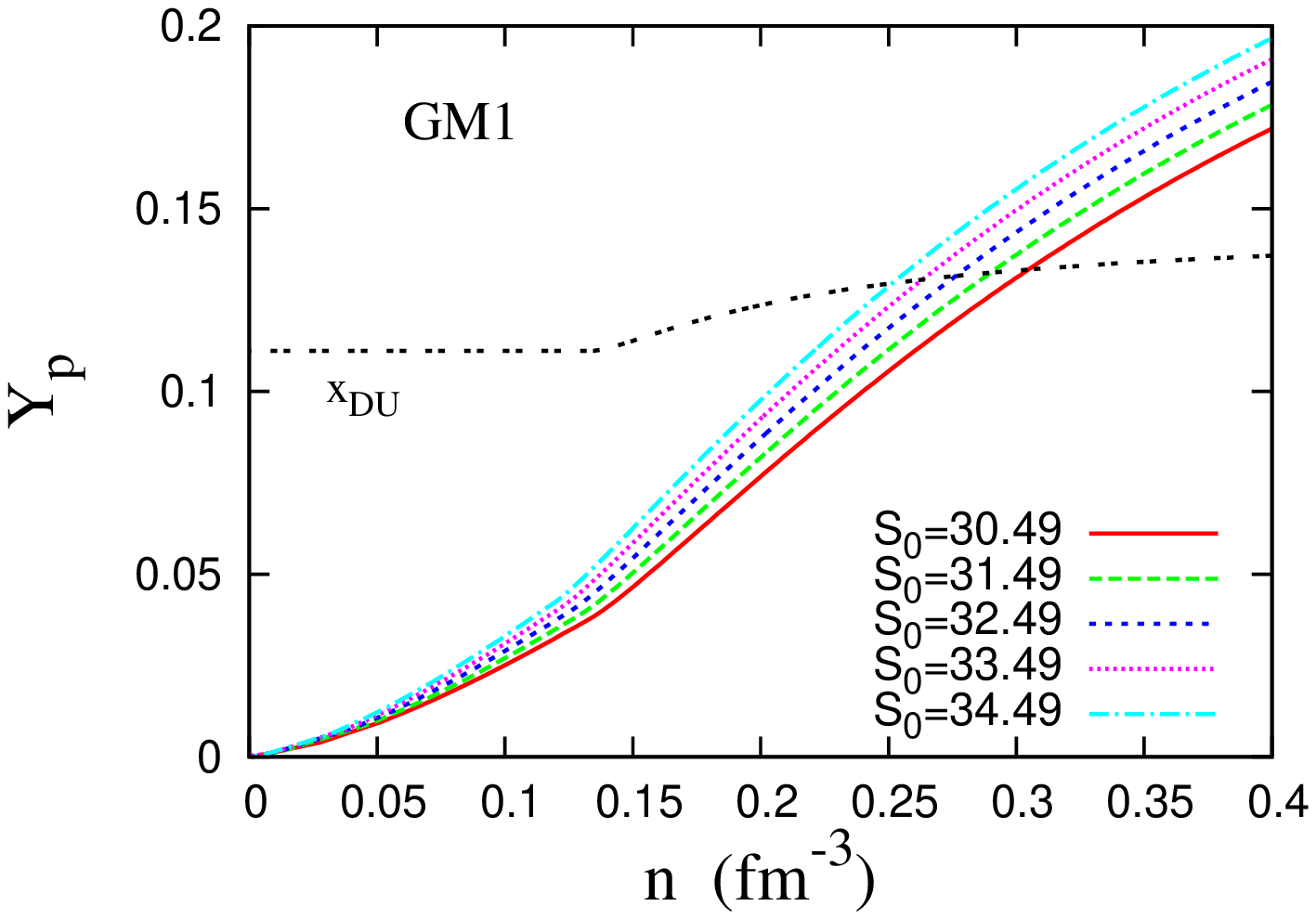} &
\includegraphics[width=5.6cm,height=6.2cm,angle=270]{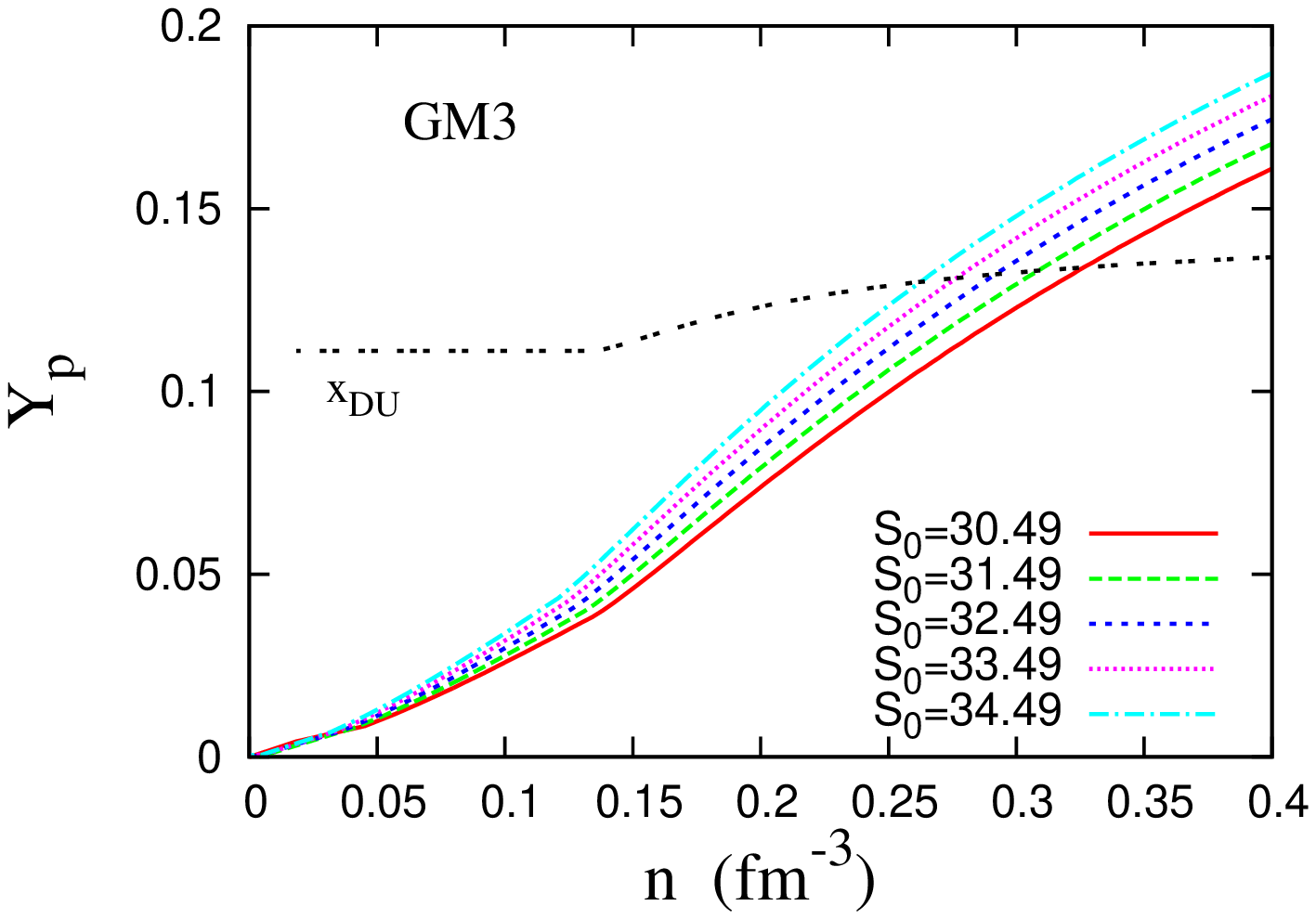} \\
\includegraphics[width=5.6cm,height=6.2cm,angle=270]{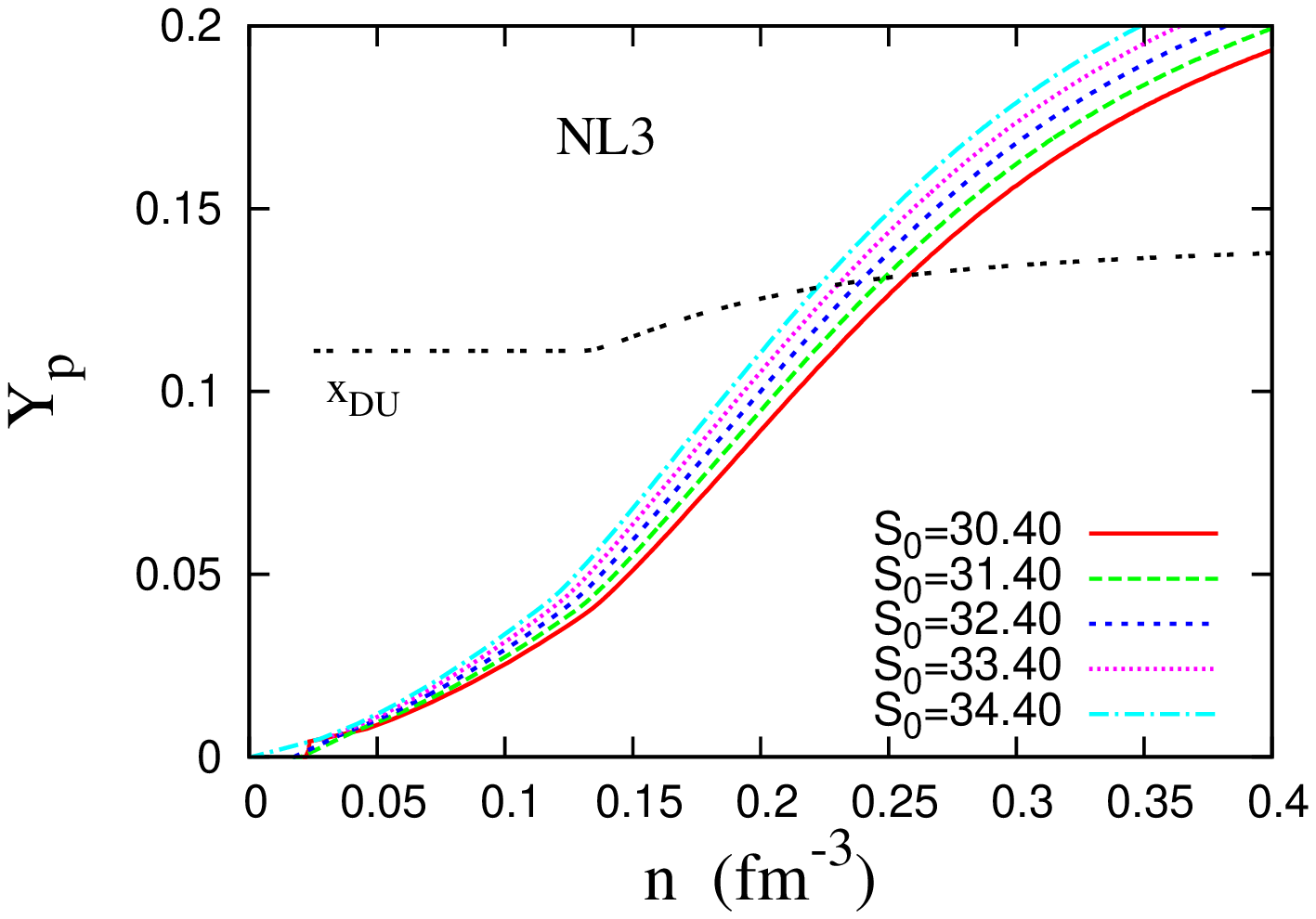} &
\includegraphics[width=5.6cm,height=6.2cm,angle=270]{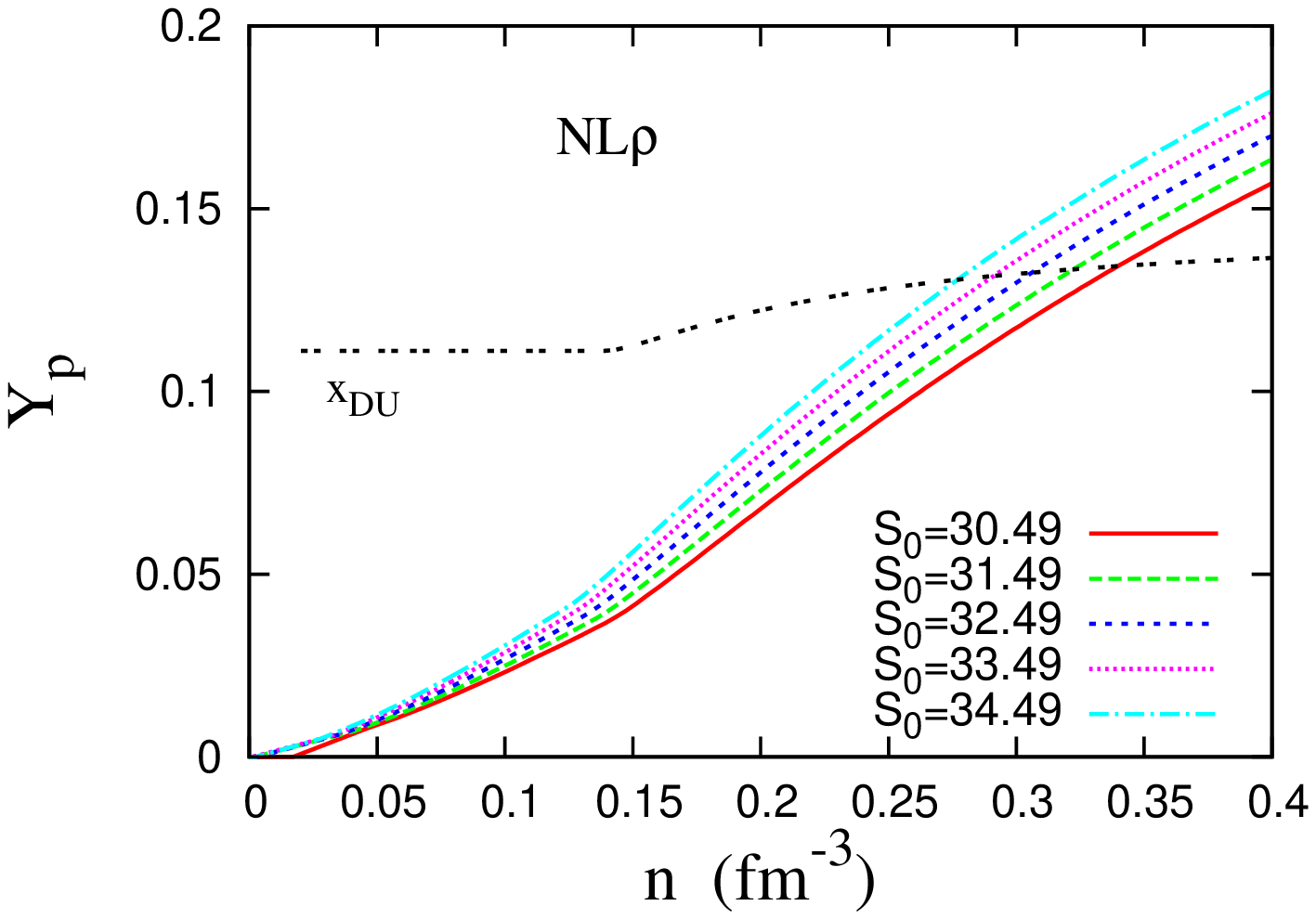} \\
\end{tabular}
\caption{(Color online) Proton fraction $Y_p$ and  the critical  value $x_{DU}$
 without $\delta$ meson. } \label{F4}
\end{figure*}

\begin{figure}[ht] 
\begin{centering}
 \includegraphics[angle=270,
width=0.4\textwidth]{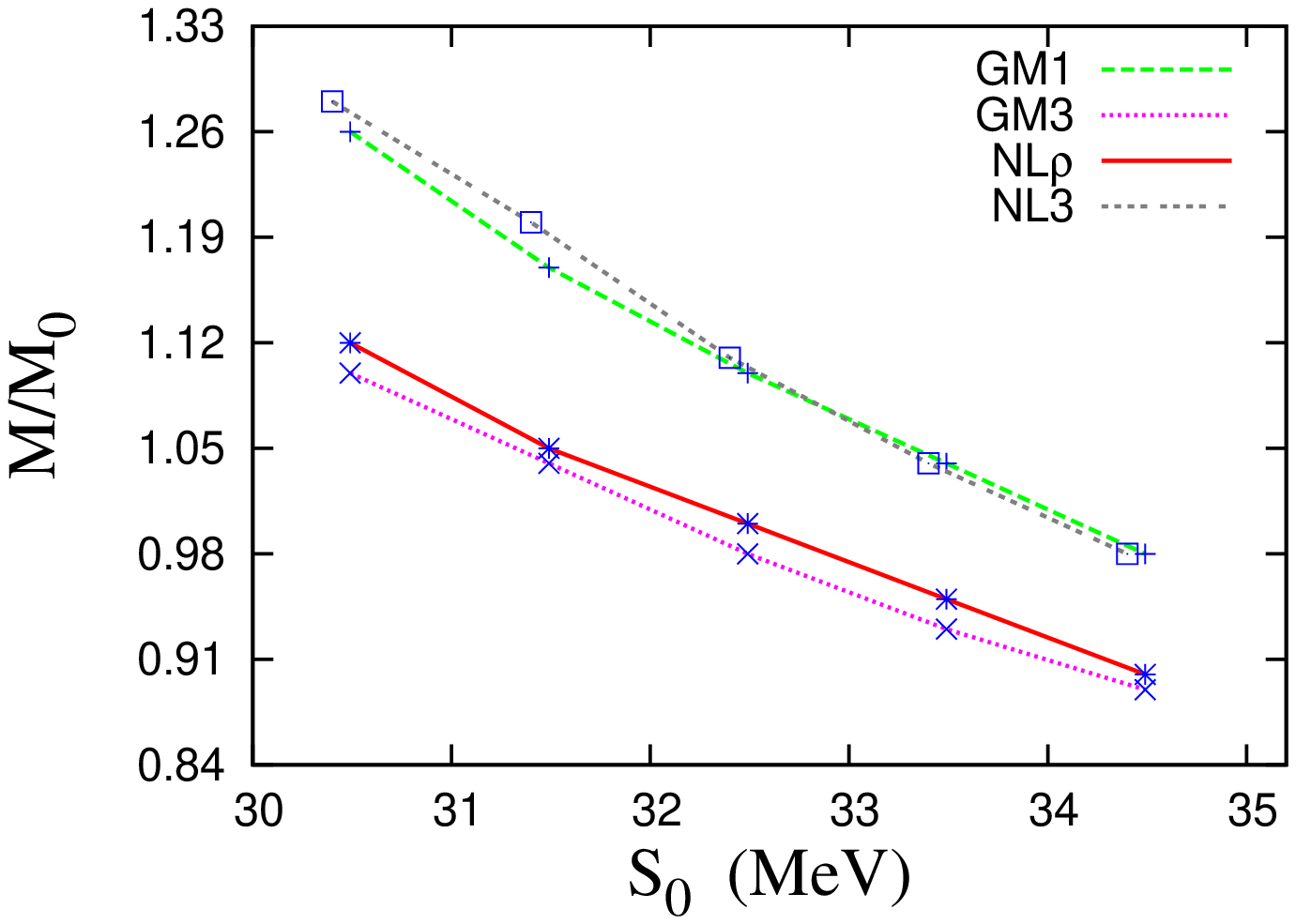}
\caption{(Color online) Minimum mass that enable DU process without $\delta$ meson} \label{F5}
\end{centering}
\end{figure}

Lets us turn now to the direct URCA process. Cooling of the neutron star by neutrino emission  can
occur much faster if direct URCA process  is allowed~\cite{Lattimer}. The direct URCA (DU) process
takes place when the proton fraction exceeds a critical value $x_{DU}$, which can
be evaluated in terms of the leptonic fraction~\cite{Lattimer,TK}:

\begin{equation}
x_{DU} = \frac{1}{1 + (1 + x_e^{1/3})^{3}} , \label{s25}
\end{equation}
where $x_e=n_e/(n_e+n_\mu)$, and $n_e$, $n_\mu$ are the number densities of the electron and the muon
respectively.  We plot the proton fractions and the corresponding neutron star masses that allow 
DU process for the four parametrizations in Figs. \ref{F4} and \ref{F5} respectively.
Although the macroscopic properties of the neutron stars { suffer}
almost no influence from the symmetry energy and its slope,
the minimum mass that enables DU process is strongly affected by them,
and could vary  up to 25 $\%$, reaching $0.3M_\odot$.

The possibility of DU process in neutron star interiors is a subject
of several  studies~\cite{Rafa2011,TK,Tsuruta,Pieka,Yakov,Than} and
some ambiguities are still present. While non relativistic models
predict a minimum mass of 1.35$M_\odot$~\cite{TK} or even higher~\cite{Tsuruta}
to allow the DU process to occur, relativistic models indicate that the minimum 
mass is as low as 0.8$M_\odot$~\cite{Pieka}. Although there is no
consensus about the minimum mass that is able to trigger the DU process, it is 
reasonable to assume  1.1$M_\odot$ as an inferior limit~\cite{Yakov}.
In this case, almost all the  parametrizations without $\delta$ meson
 should be avoided, as can be seen from  by Fig. \ref{F5}.

We resume the main results  of this section in Table \ref{T14}.
In general, the parametrizations that predict higher maximum masses
also predict larger radii for the canonical mass and higher minimum mass
 that enables DU process. However, this is not a general rule, since NL$\rho$
predicts a higher mass than GM3 (2.11 $M_\odot$ vs 2.04 $M_\odot$) even with 
 a lower radius  value for the $1.4 M_\odot$ (12.93 km vs 13.07 km).
Also, one can  notice that the NL$\rho$  has a bigger slope than GM3 
(85.0 MeV vs 83.7 MeV with the $S_0 = 30.49$), indicating that the 
knowledge of the slope $L$  is not enough to infer the neutron
star radius.

\begin{table}[ht]
\begin{center}
\begin{tabular}{|c|c|c|c|c|c|}
\hline 
 Model &  $S_0$ (MeV) & $M_{max}/M_\odot$  & $R_{1.4M_\odot}$ & $M_{DU}/M_\odot$ & $n_{DU}$ ($fm^{-3}$)   \\
 \hline
 GM1 & 30.49  & 2.39 & 13.72 & 1.26 & 0.305   \\
 \hline
GM1 & 31.49 & 2.39 & 13.76 & 1.17 &  0.290 \\
 \hline
 GM1 & 32.49 & 2.39 & 13.80 & 1.10 & 0.279 \\
\hline
 GM1 & 33.49 & 2.38 & 13.84 & 1.04 & 0.267  \\
\hline  
 GM1 & 34.49 & 2.38 & 13.91 & 0.98 & 0.255  \\
\hline
\hline
GM3 & 30.49 & 2.04 & 13.07 & 1.10 & 0.327 \\
\hline
GM3 & 31.49 & 2.04 & 13.12 & 1.04 & 0.309 \\
\hline
GM3 & 32.49 & 2.04 & 13.16 & 0.98 & 0.293 \\
\hline
GM3 & 33.49 & 2.04 & 13.22 & 0.93 & 0.280 \\
\hline
GM3 & 34.49 & 2.04 & 13.26 & 0.89 & 0.267 \\
\hline
\hline
NL3 & 30.40 & 2.81 & 14.44 & 1.28 & 0.258 \\
\hline
NL3 & 31.41 & 2.81 & 14.48 & 1.20 & 0.249 \\
\hline
NL3 & 32.40 & 2.80 & 14.51 & 1.11 & 0.241  \\
\hline
NL3 & 33.40 & 2.80 & 14.54 & 1.04 & 0.232 \\
\hline
NL3 & 34.40 & 2.79 & 14.61 & 0.98 & 0.225 \\
\hline
\hline
NL$\rho$  & 30.49 & 2.11 & 12.93 & 1.12 & 0.340 \\
\hline
NL$\rho$ & 31.49 & 2.11 & 12.97 & 1.05 & 0.323 \\
\hline
NL$\rho$ & 32.49 & 2.11 & 13.01 & 1.00 & 0.308 \\
\hline
NL$\rho$  & 33.49 & 2.11 & 13.07 & 0.95 & 0.295 \\
\hline
NL$\rho$ & 34.49 & 2.10 & 13.13 & 0.90 & 0.279 \\
\hline
\end{tabular} 
\caption{Neutron star main properties without $\delta$ meson.} 
\label{T14}
\end{center}
\end{table}


\subsection{Fixing the slope $L$}

In a second attempt,  with the 
{ inclusion} of the $\delta$ meson,
 we fix the slope and vary the  symmetry energy
 to study the influence of this quantity 
independently. The parameters utilized in this approach
 are presented in Tables \ref{T6}, \ref{T7}, \ref{T8} and \ref{T9}
for GM1,  GM3,  NL3 and  NL$\rho$ respectively.

\begin{table}[!htb]
\begin{center}
\begin{tabular}{|c|c||c|c|}
\hline 
$(g_\rho/m_\rho)^2$ ($fm^{-2}$)  &  $(g_\delta/m_\delta)^2$ ($fm^{-2}$) &  $S_0$ (MeV)  & L (MeV)   \\
 \hline
 9.577 & 1.68  & 30.16 & 99.9   \\
 \hline
 8.435 & 1.26 & 31.28 & 99.9 \\
 \hline
 7.282 & 0.84 & 32.38 & 99.9  \\
\hline
 6.120 & 0.42 & 33.46 & 99.9  \\
\hline  
 4.936 & 0.00 & 34.49 & 99.9  \\
\hline
\end{tabular} 
\caption{ $S_0$ values  with GM1 for a fixed slope  at 99.9 MeV} 
\label{T6}
\end{center}
\end{table}

\begin{table}[!htb]
\begin{center}
\begin{tabular}{|c|c||c|c|}
\hline 
$(g_\rho/m_\rho)^2$ ($fm^{-2}$)  &  $(g_\delta/m_\delta)^2$ ($fm^{-2}$) &  $S_0$ (MeV)  & L (MeV)   \\
 \hline
 12.521 & 2.40  & 30.15 & 95.7   \\
 \hline
 10.850 & 1.80 & 31.32 & 95.7 \\
 \hline
 9.025 & 1.20 & 32.42 & 95.7  \\
\hline
 7.183 & 0.60 & 33.48 & 95.7  \\
\hline  
 5.319 & 0.00 & 34.49 & 95.7  \\
\hline
\end{tabular} 
\caption{$S_0$ values  with GM3 for a fixed slope   at 95.7 MeV} 
\label{T7}
\end{center}
\end{table}

\begin{table}[!htb]
\begin{center}
\begin{tabular}{|c|c||c|c|}
\hline 
$(g_\rho/m_\rho)^2$ ($fm^{-2}$)  &  $(g_\delta/m_\delta)^2$ ($fm^{-2}$) &  $S_0$ (MeV)  & L (MeV)   \\
 \hline
 6.80 & 1.04  & 30.11 & 109.4   \\
 \hline
 6.246 & 0.78 & 31.19 & 109.4 \\
 \hline
 5.690 & 0.52 & 32.28 & 109.4  \\
\hline
 5.127 & 0.26 & 33.36 & 109.4  \\
\hline  
 4.552 & 0.00 & 34.40 & 109.4  \\
\hline
\end{tabular} 
\caption{$S_0$ values  with NL3 for a fixed slope  at 109.4 MeV} 
\label{T8}
\end{center}
\end{table}

\begin{table}[!htb]
\begin{center}
\begin{tabular}{|c|c||c|c|}
\hline 
$(g_\rho/m_\rho)^2$ ($fm^{-2}$)  &  $(g_\delta/m_\delta)^2$ ($fm^{-2}$) &  $S_0$ (MeV)  & L (MeV)   \\
 \hline
 10.696 & 2.00  & 30.04 & 97.0   \\
 \hline
 9.260 & 1.50 & 31.22 & 97.0 \\
 \hline
 7.803 & 1.00 & 32.35 & 97.0  \\
\hline
 6.333 & 0.50 & 33.45 & 97.0  \\
\hline  
 4.842 & 0.00 & 34.49 & 97.0  \\
\hline
\end{tabular} 
\caption{$S_0$ values  with NL$\rho$ for a fixed slope  at 97.0 MeV} 
\label{T9}
\end{center}
\end{table}

\begin{figure*}[hb]
\begin{tabular}{cc}
\includegraphics[width=5.6cm,height=6.2cm,angle=270]{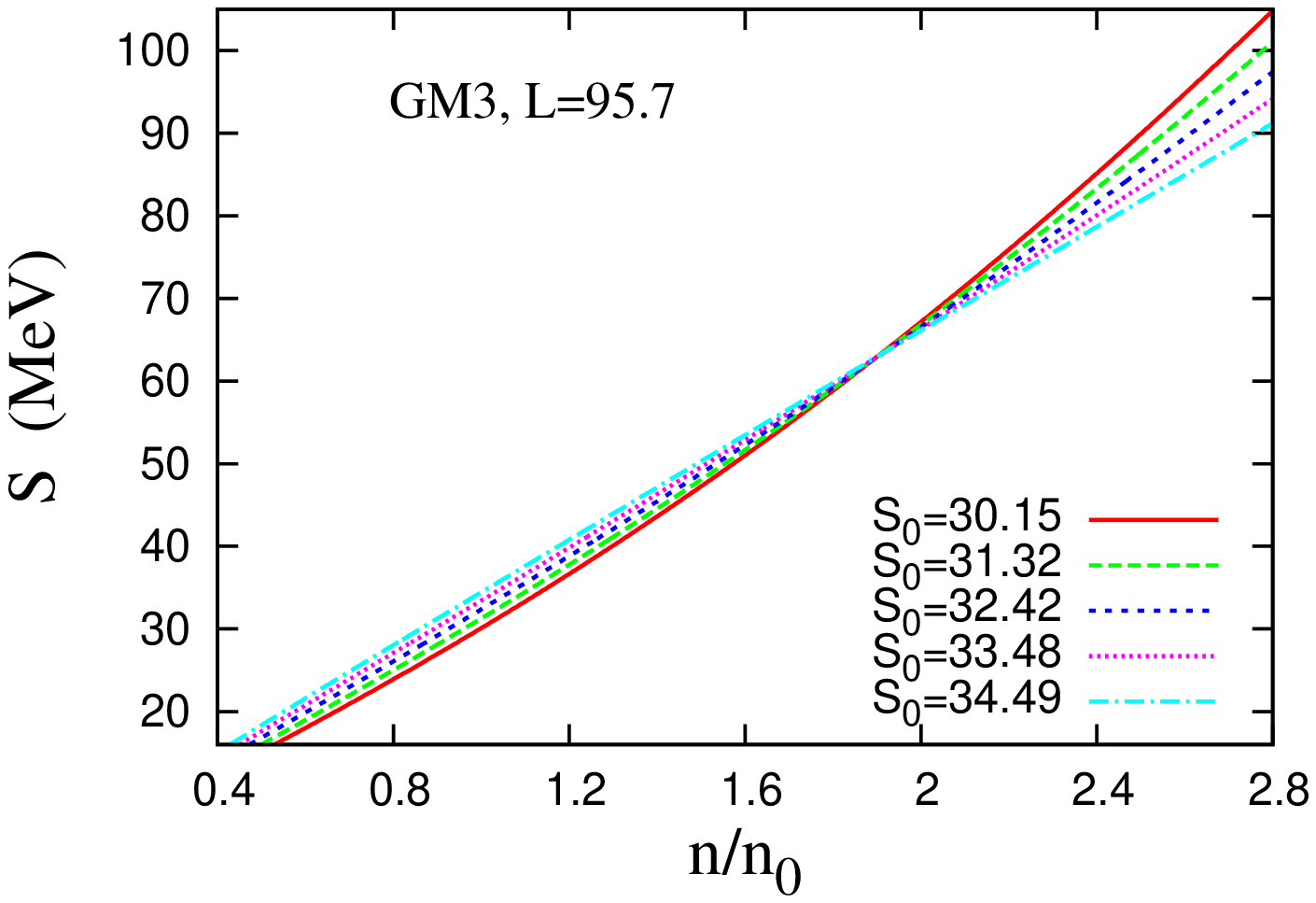} &
\includegraphics[width=5.6cm,height=6.2cm,angle=270]{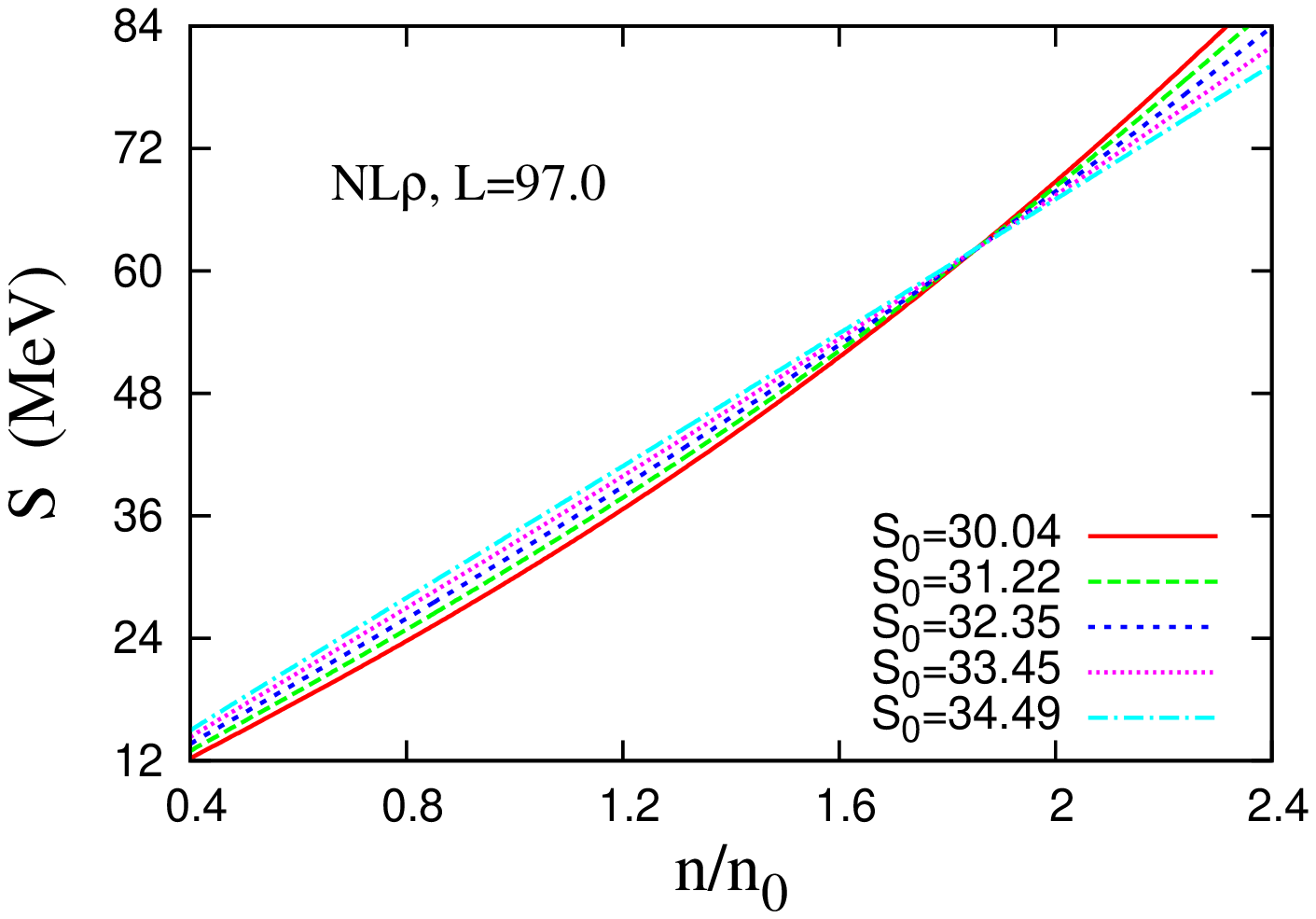} \\
\includegraphics[width=5.6cm,height=6.2cm,angle=270]{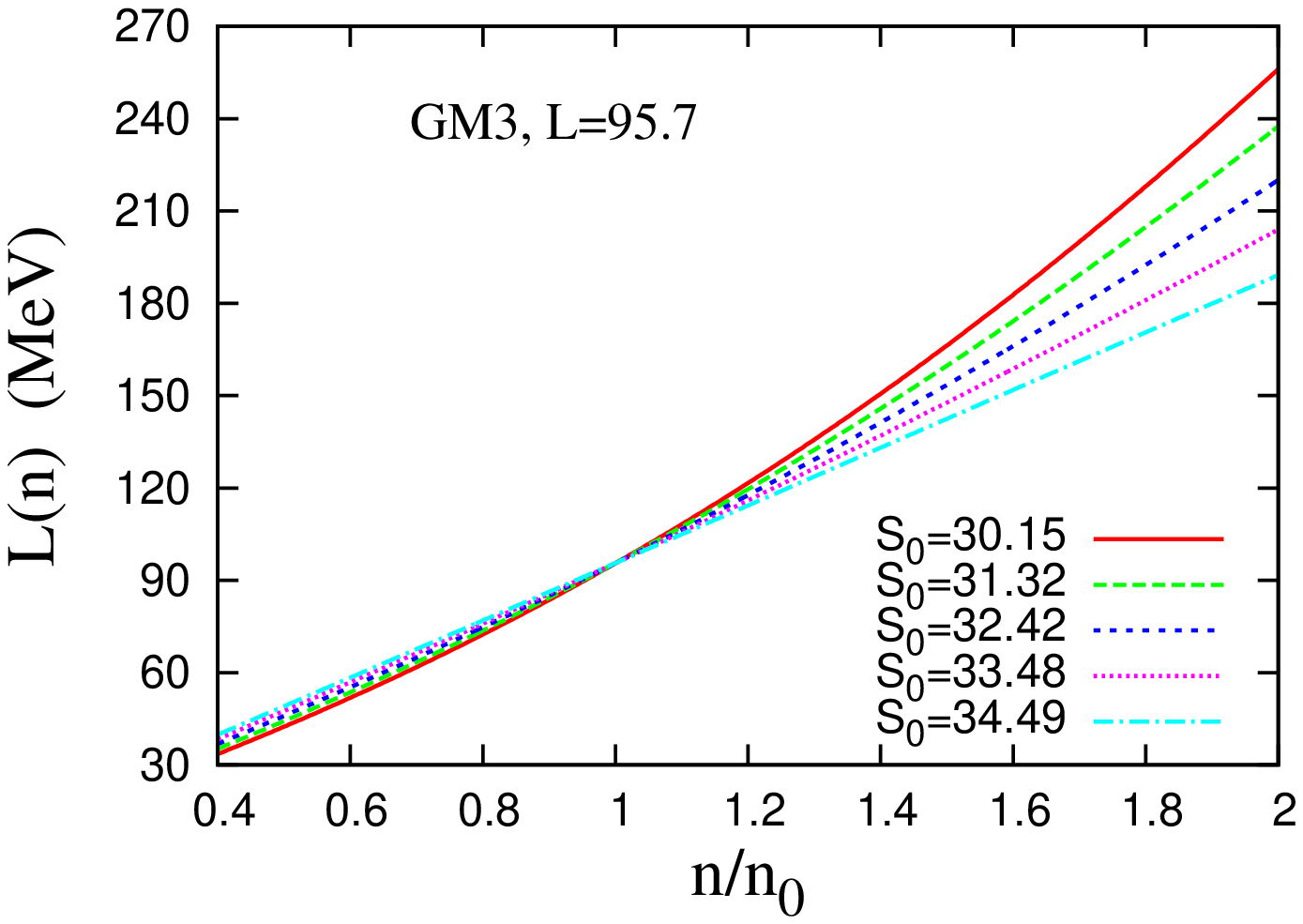} &
\includegraphics[width=5.6cm,height=6.2cm,angle=270]{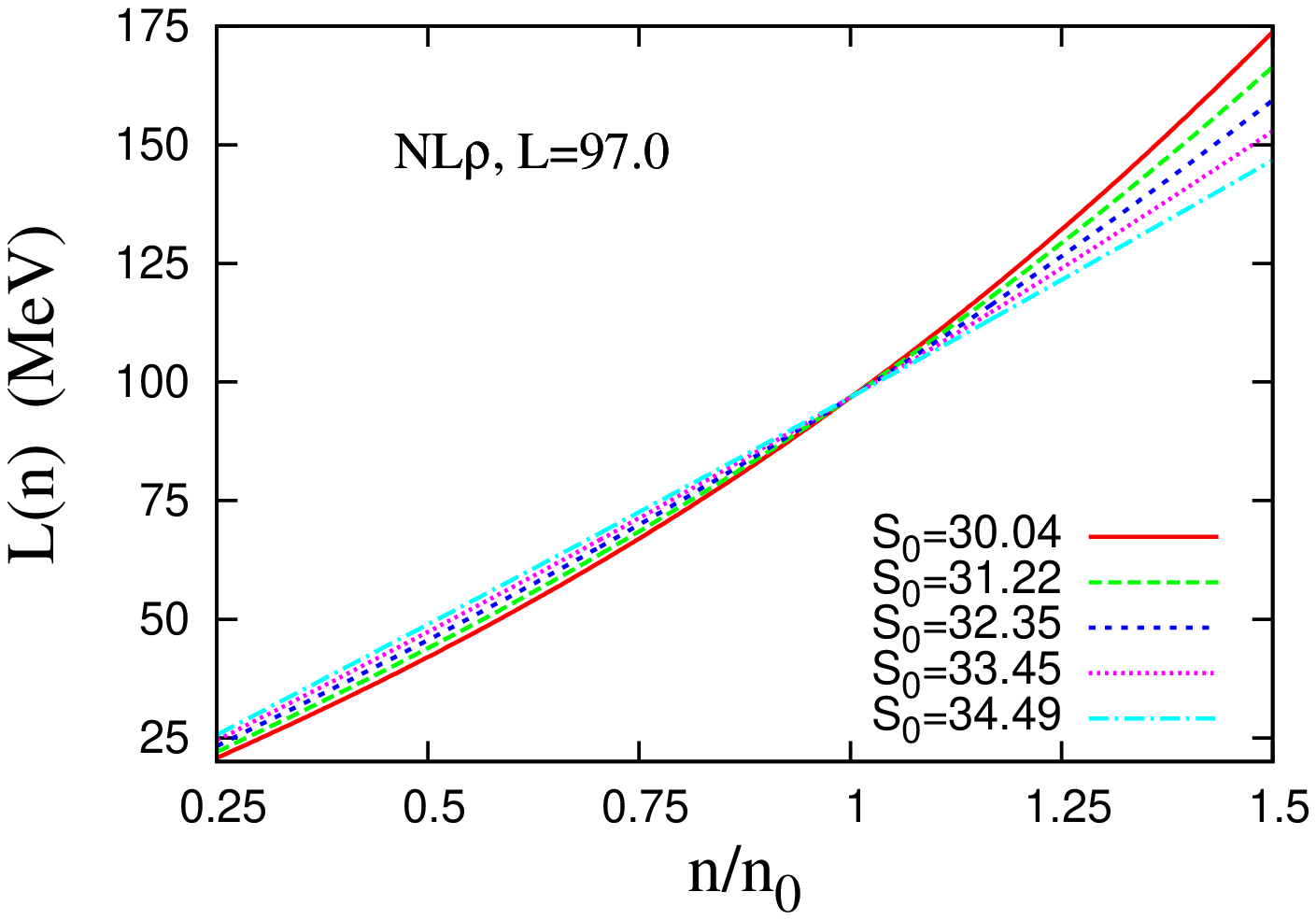} \\
\end{tabular}
\caption{(Color online) (Top) Symmetry energy $S$ as function of density and (Bottom) slope of the 
symmetry energy $L(n)$ as function of density with fixed $L$. } \label{F6}
\end{figure*}

 We plot the density dependent symmetry energy $S$ and the the slope $L(n)$ with fixed $L$
for GM3 and NL$\rho$ parametrization in Fig. \ref{F6}.

When we fix the slope,  the $S$ and  $L(n)$ curves always cross each other.
 The parametrizations that predict 
lower values of $S$ and $L(n)$ at low density always  predict high values at higher densities, contrary
to what was found in the previous section. This is a global effect, present in our four models.
The reason is that alongside the repulsive $\rho$ meson, the attractive $\delta$ meson contributes both to the symmetry energy 
and  the slope. The $\delta$ meson, being scalar,  dominates at low densities, 
while the $\rho$ meson, being a vector meson,  dominates at high densities. So, to fix the slope at certain
value, the lower the $S$ and the $L(n)$ are at low densities, the higher they are at high density.

 \begin{figure*}[hb]
\begin{tabular}{cc}
\includegraphics[width=5.6cm,height=6.2cm,angle=270]{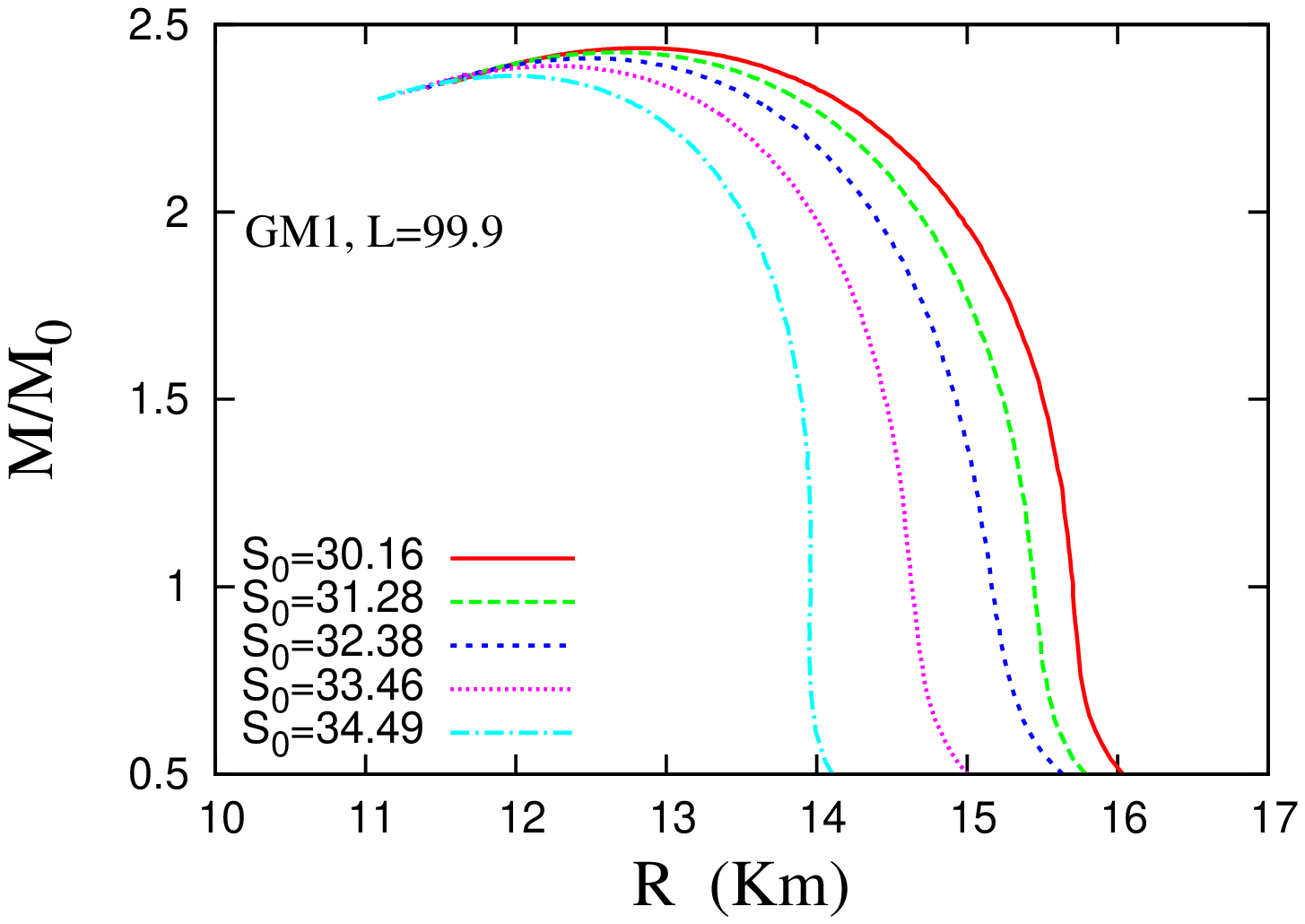} &
\includegraphics[width=5.6cm,height=6.2cm,angle=270]{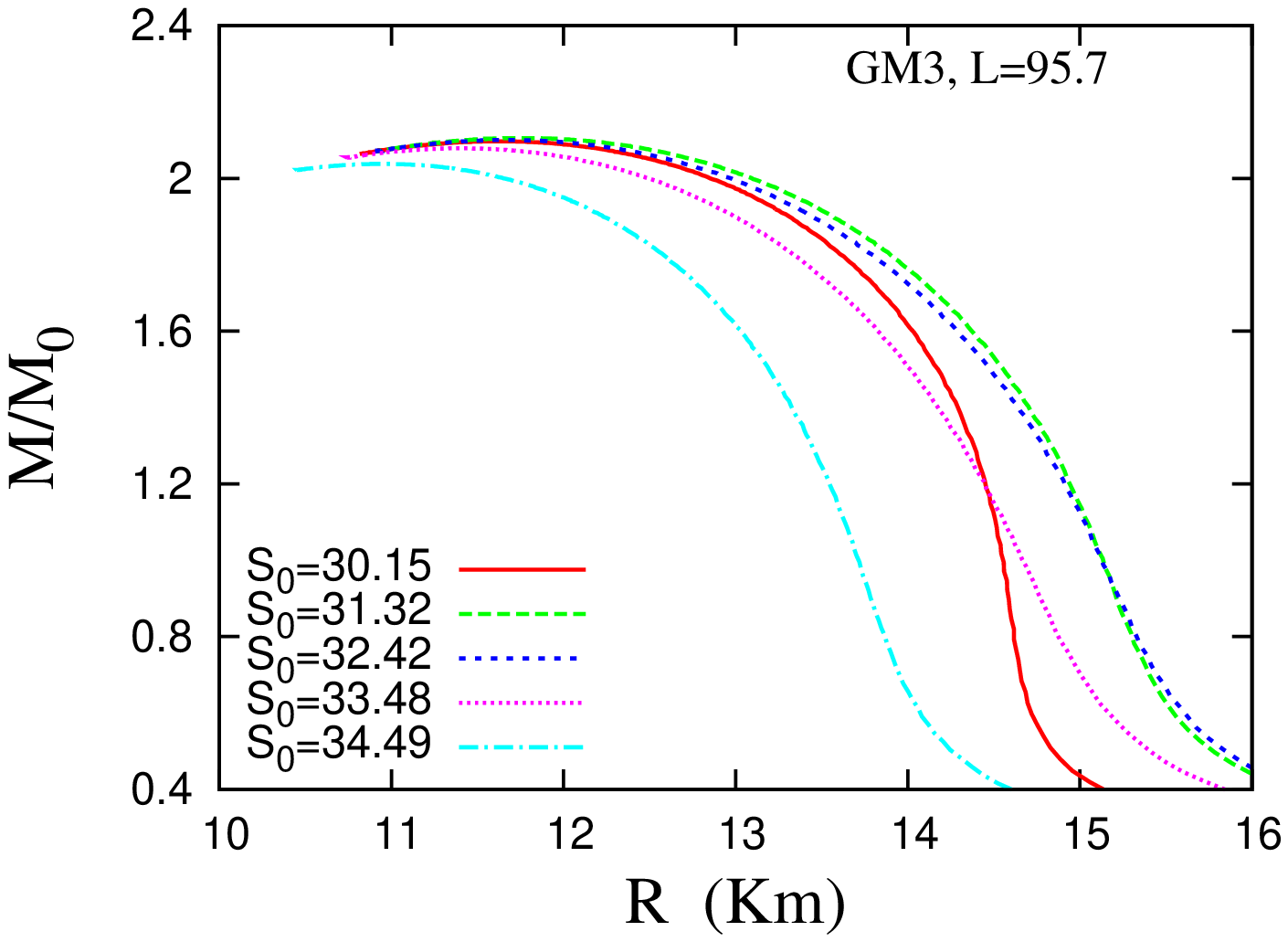} \\
\includegraphics[width=5.6cm,height=6.2cm,angle=270]{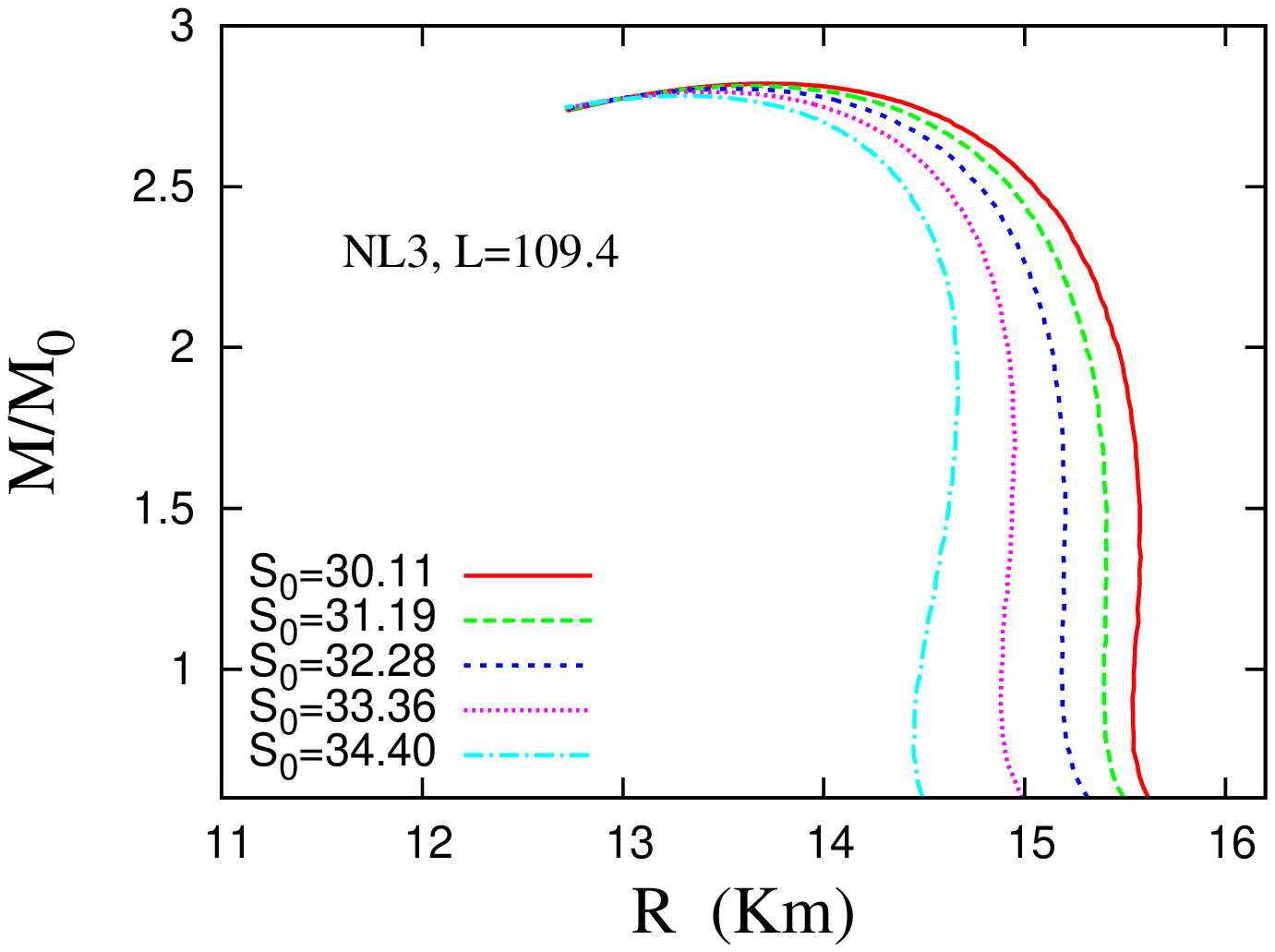} &
\includegraphics[width=5.6cm,height=6.2cm,angle=270]{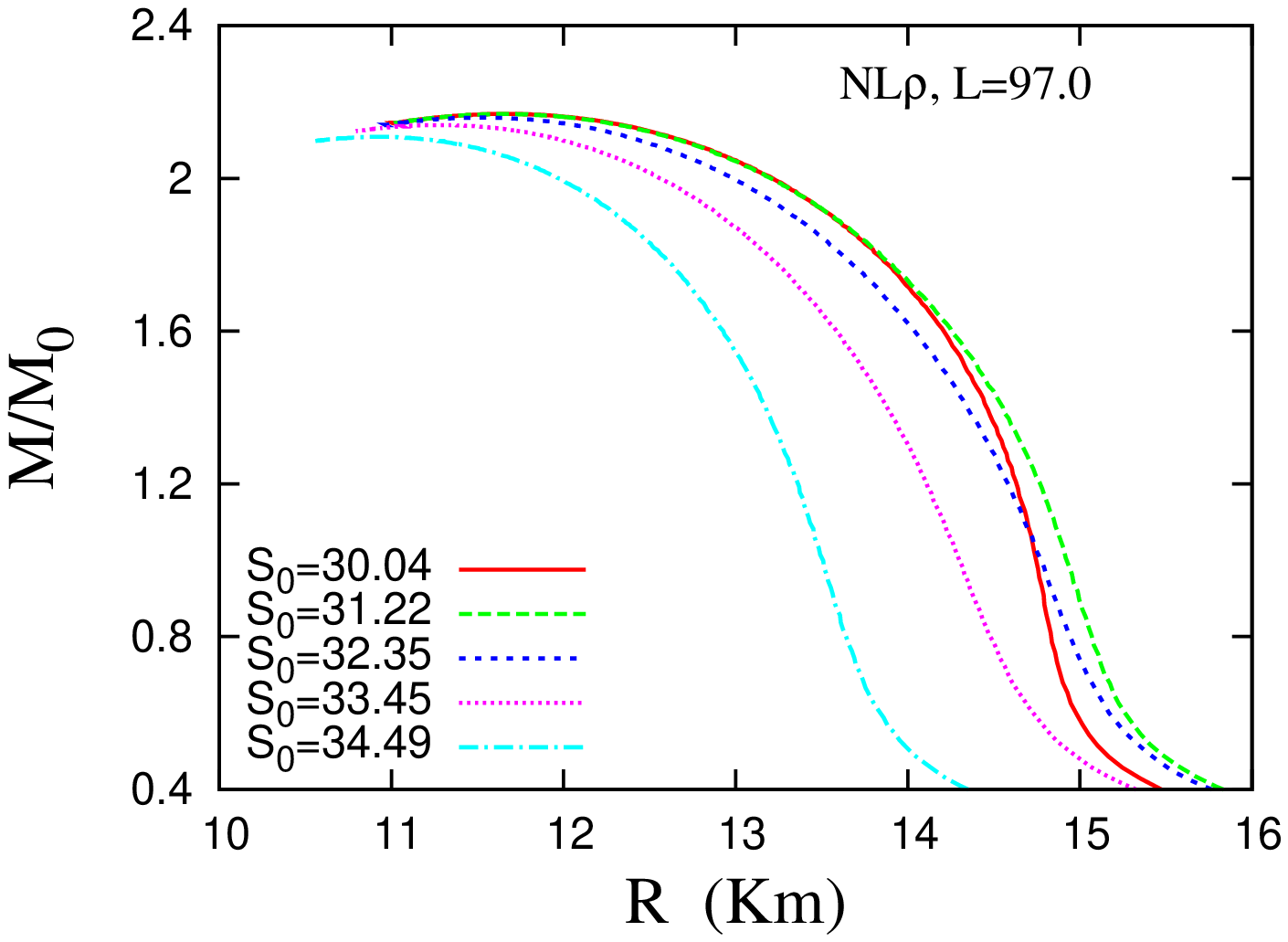} \\
\end{tabular}
\caption{(Color online) Neutron star mass-radius relation with fixed $L$. } \label{F8}
\end{figure*}

Now we study this crossing effect in neutron star properties. We plot the mass/radius relation in 
Fig. \ref{F8}. When we fix the slope $L$, a curious behaviour appears. The radii of
canonical 1.4$M_\odot$ in general decreases with the symmetry energy.
This effect, as far as we know, { has not been} noticed before.
Usually, when $\delta$ meson is not included, the $S_0$ is determined
just by the $\rho$ meson coupling, and its slope is uniquely obtained.
A correlation between $L$ and the neutron star radius is then seen~\cite{Rafa2011}.
When we fix the slope, with the help of the $\delta$ meson, different values of $S_0$
are then obtained, and the correlation is lost.
 So, although the slope $L$ give us significant 
{ information on the behaviour of the masses}, it is not
enough to determine the neutron star radius, since different parametrizations
with  the same value of $L$ cause variations of up to  1.7 km.

We can also see, that for GM3 and NL$\rho$, as $S_0$ increases
the radius first increases a little, then strongly decreases.
 This behaviour could indicate that, in these parametrizations,  the  value
utilized for  the slope $L$ (95.7 for GM3 and 97.0 for NL$\rho$)
 is too large  for a symmetry energy $S_0$ around 30 MeV.

The maximum mass changes a little more than in the case without $\delta$
meson, but it is still not significantly altered,
 being not superior to $0.05M_\odot$.

\begin{figure*}[hb]
\begin{tabular}{cc}
\includegraphics[width=5.6cm,height=6.2cm,angle=270]{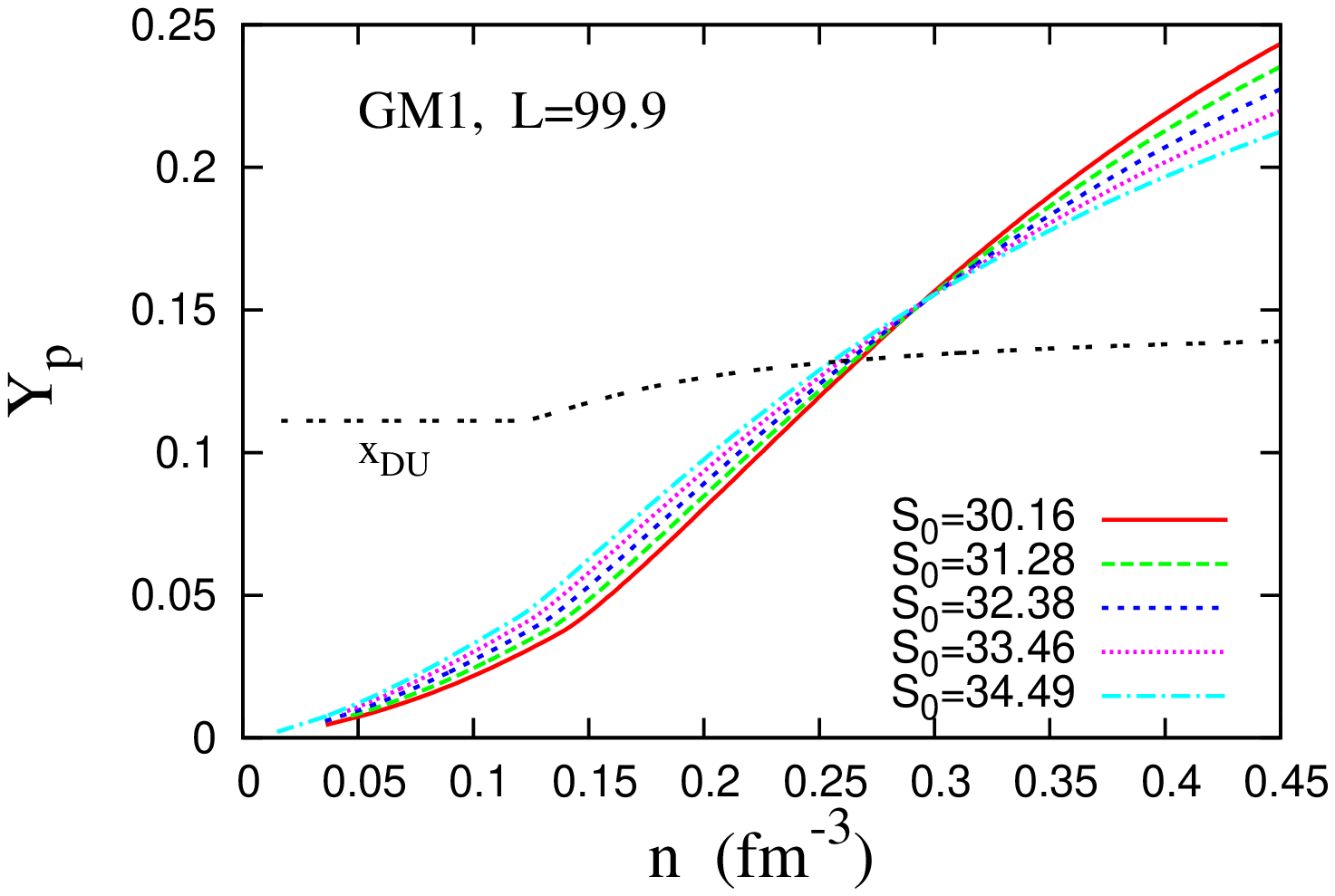} &
\includegraphics[width=5.6cm,height=6.2cm,angle=270]{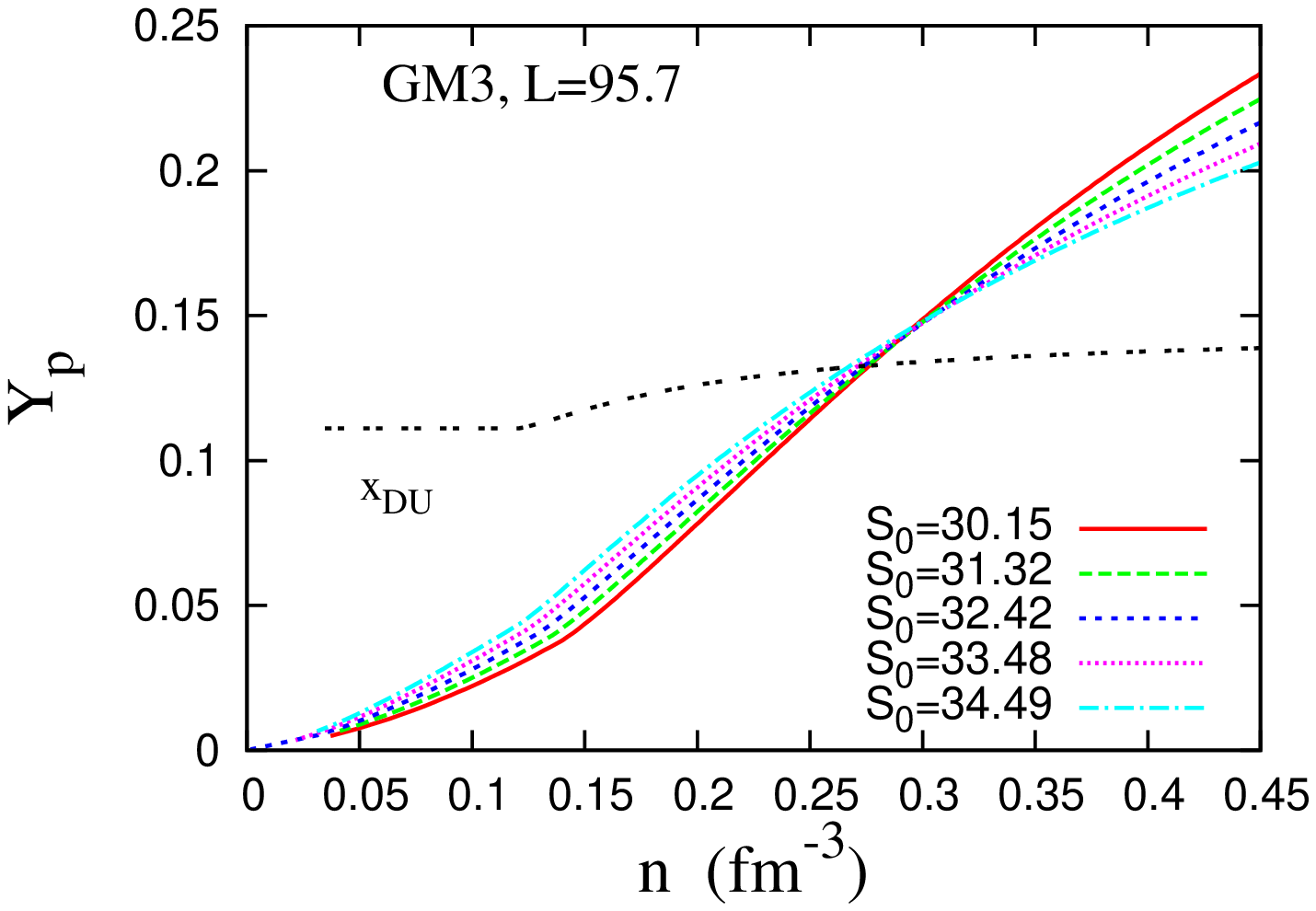} \\
\includegraphics[width=5.6cm,height=6.2cm,angle=270]{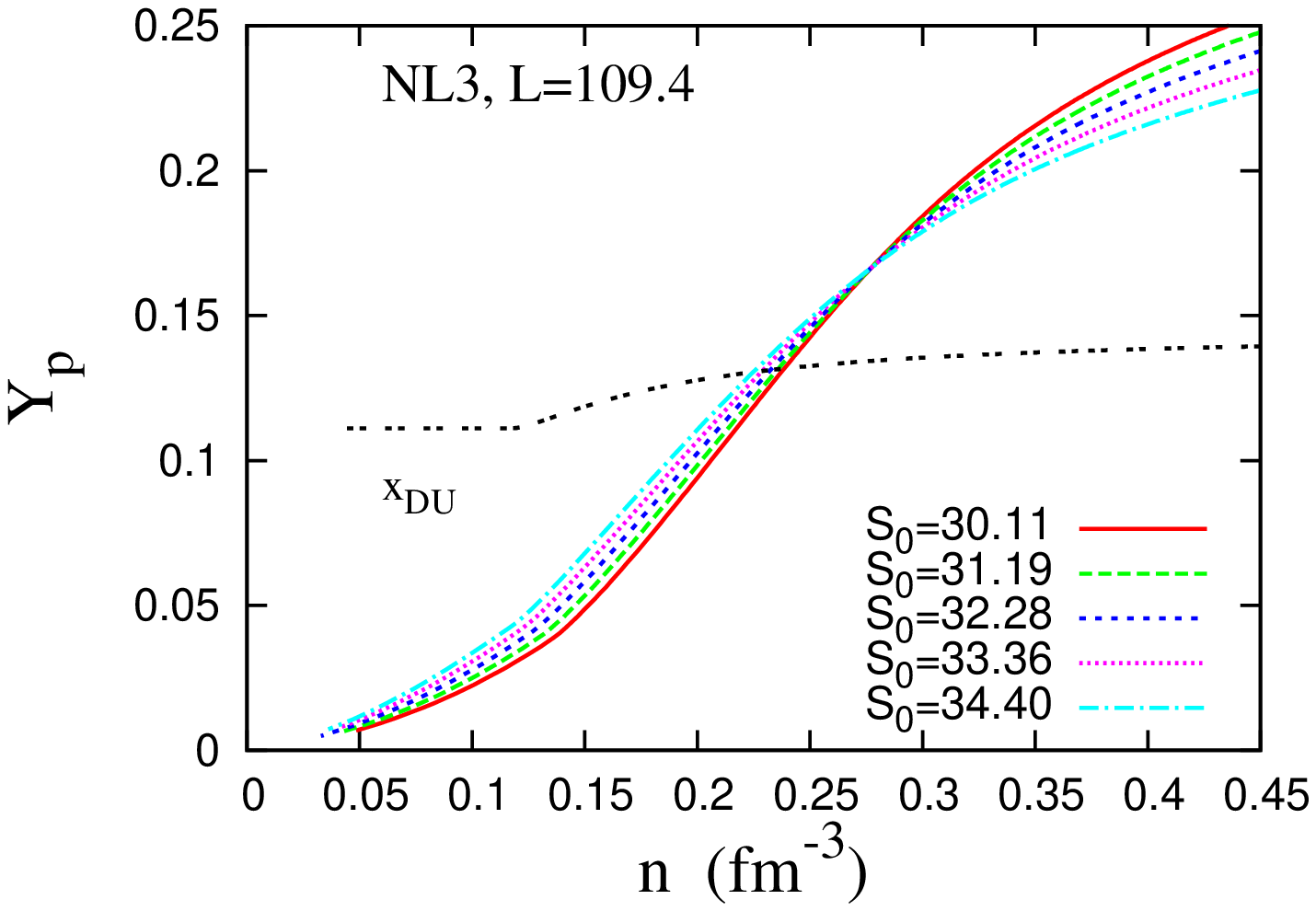} &
\includegraphics[width=5.6cm,height=6.2cm,angle=270]{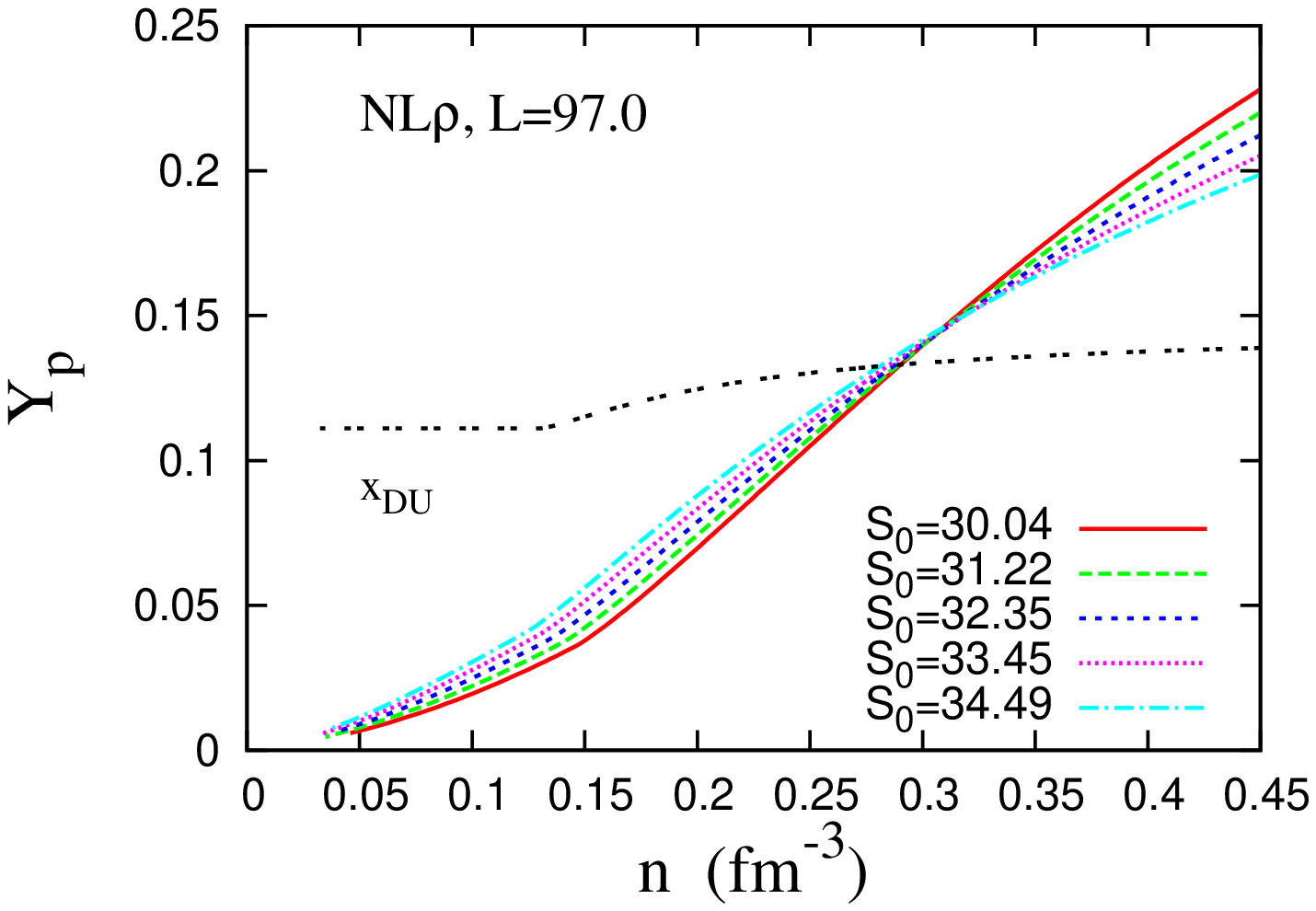} \\
\end{tabular}
\caption{(Color online) Proton fraction $Y_p$ and  the critical  value $x_{DU}$
 with fixed $L$.  } \label{F9}
\end{figure*}

To study the DU process, we plot the proton fraction in Fig. \ref{F9}.
We see that the same behaviour present in the symmetry energy $S$ and
in the slope $L(n)$ is observed in the proton fraction $Y_p$, i.e.
the parametrizations with lower values of $Y_p$ at low densities
are those with higher values of it at high densities.
The reason  again lies in the competition between the 
scalar $\delta$ meson and the vector $\rho$ meson.
The $\delta$ meson dominates at low densities, reducing the mass
of the neutrons and increasing the mass of the protons, favouring
neutron population. However, at high densities the $\rho$ meson
dominates, reducing the proton chemical potential. 
This inversion of lower/higher proton fraction in all models
happens for densities above the critical value of $x_{DU}$.

\begin{figure}[hb] 
\begin{centering}
 \includegraphics[angle=270,
width=0.4\textwidth]{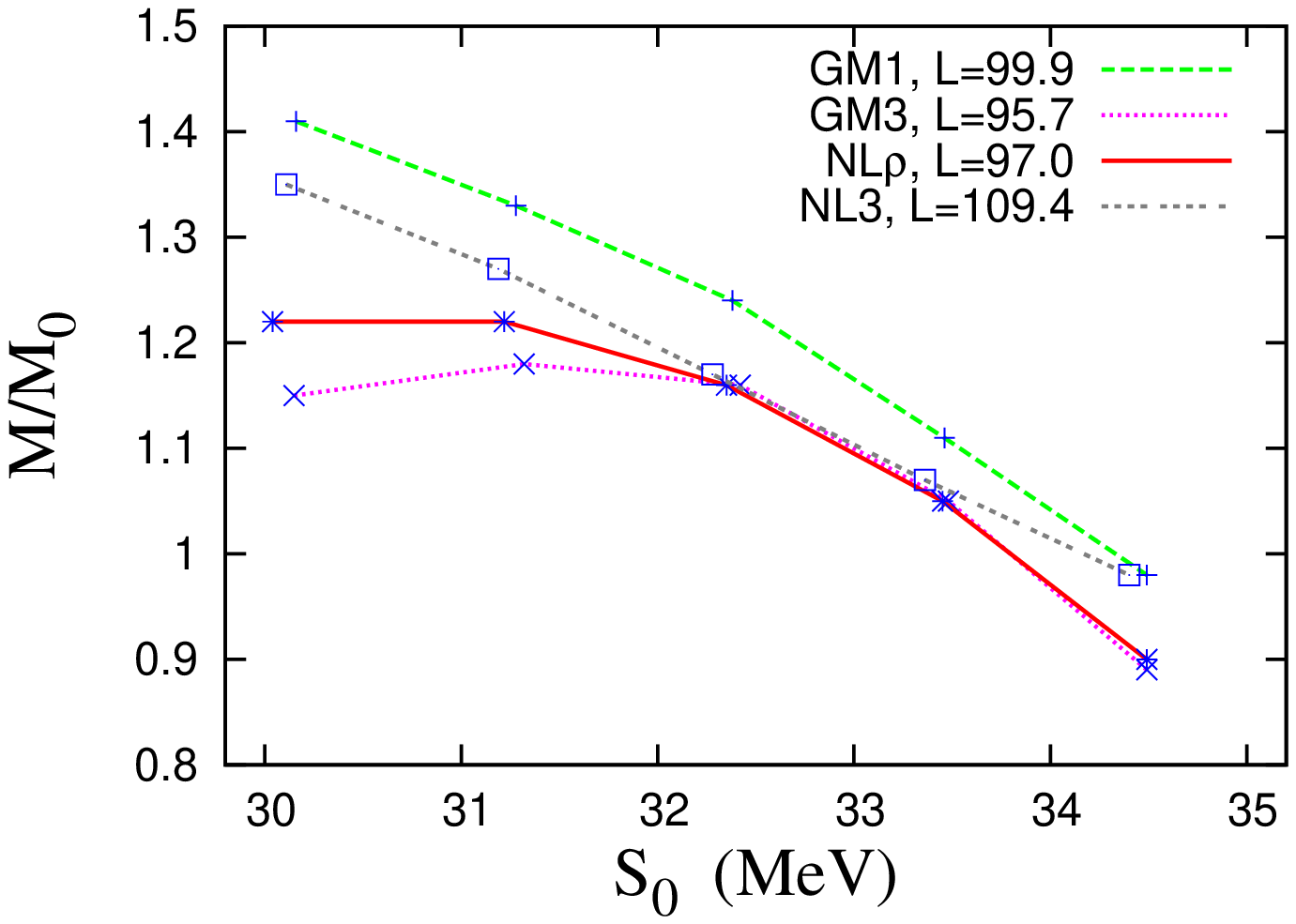}
\caption{(Color online) Minimum mass that enable DU process with fixed $L$.} \label{F10}
\end{centering}
\end{figure}

The corresponding neutron star mass that allows the
DU process is shown in Fig. \ref{F10}. We see that, in general, increasing the symmetry energy $S_0$
leads to a reduction of the minimum mass that enables the DU process, as in the case without $\delta$
meson. It implies that, if on one hand the symmetry energy $S_0$ has little
influence on the radius of the canonical $1.4 M_\odot$ neutron star, on the other hand, it
has a strong influence on the minimum mass that enables DU process, (although
the slope $L$  still contributes). We see that even different parametrizations
result in  very similar masses with similar values of $S_0$. Also, it  is worth  noting that
when the symmetry energy increases from  $\approx$ 30 to $\approx$ 31 MeV, the
minimum masses for GM3 and NL$\rho$ also increase. This effect is similar
to the one present  in the radius/symmetry energy relation,
and again could indicate that the values of  $L$   are too large for $S_0$  around 30 MeV.

We resume  this section in Table \ref{T15}.

\begin{table}[!htb]
\begin{center}
\begin{tabular}{|c|c|c|c|c|c|}
\hline 
 Model &  $S_0$ (MeV) & $M_{max}/M_\odot$  & $R_{1.4M_\odot}$ & $M_{DU}/M_\odot$ & $n_{DU}$ ($fm^{-3}$)   \\
 \hline
 GM1 & 30.16  & 2.44 & 15.56 & 1.41 & 0.266   \\
 \hline
GM1 & 31.28 & 2.42 & 15.29 & 1.33 &  0.265 \\
 \hline
 GM1 & 32.38 & 2.41 & 14.98 & 1.24 & 0.262 \\
\hline
 GM1 & 33.46 & 2.39 & 14.50 & 1.10 & 0.258  \\
\hline  
 GM1 & 34.49 & 2.38 & 13.91 & 0.98 & 0.255  \\
\hline
\hline
GM3 & 30.15 & 2.09 & 14.29 & 1.15 & 0.277 \\
\hline
GM3 & 31.32 & 2.09 & 14.70 & 1.18 & 0.275 \\
\hline
GM3 & 32.42 & 2.09 & 14.63 & 1.16 & 0.274 \\
\hline
GM3 & 33.48 & 2.07 & 14.17 & 1.05 & 0.271 \\
\hline
GM3 & 34.49 & 2.04 & 13.26 & 0.89 & 0.267 \\
\hline
\hline
NL3 & 30.11 & 2.83 & 15.57 & 1.35 & 0.237 \\
\hline
NL3 & 31.19 & 2.82 & 15.40 & 1.27 & 0.235 \\
\hline
NL3 & 32.28 & 2.81 & 14.20 & 1.17 & 0.231  \\
\hline
NL3 & 33.36 & 2.80 & 14.93 & 1.07 & 0.228 \\
\hline
NL3 & 34.40 & 2.79 & 14.61 & 0.98 & 0.225 \\
\hline
\hline
NL$\rho$  & 30.04 & 2.16 & 14.44 & 1.22 & 0.290 \\
\hline
NL$\rho$ & 31.22 & 2.16 & 14.55 & 1.23 & 0.289 \\
\hline
NL$\rho$ & 32.35 & 2.14 & 14.33 & 1.16 & 0.287 \\
\hline
NL$\rho$  & 33.45 & 2.12 & 13.88 & 1.05 & 0.284 \\
\hline
NL$\rho$ & 34.49 & 2.10 & 13.13 & 0.90 & 0.279 \\
\hline
\end{tabular} 
\caption{Neutron star main properties with fixed  $L$.} 
\label{T15}
\end{center}
\end{table}

We see that although different values of the symmetry energy
produce different minimum masses that enable the DU process, the central 
density $n$ are very similar. This is due to the fact that
the proton fractions cross each other in a density close to the $x_{DU}$.


\subsection{Fixing the symmetry energy $S_0$}

Finally, our last approach is to fix the $S_0$ value to study the 
direct influence of the  slope $L$.
The parameters obtained with this approach 
 are presented in Tables \ref{T10}, \ref{T11}, \ref{T12} and \ref{T13}
 for the GM1,  GM3,  NL3 and  NL$\rho$ respectively.
 Note that  GM1 and GM3 parametrizations keep the original value of $S_0$, fixed at
 32.49 MeV. However, for $NL\rho$ and NL3,  we change the values
 from 30.49 MeV to 31.49 MeV  and from 37.40 MeV to 33.40 MeV
 respectively, so that the the $L$ values are no more than 1 MeV apart
 from each other.

\begin{table}[!htb]
\begin{center}
\begin{tabular}{|c|c||c|c|}
\hline 
$(g_\rho/m_\rho)^2$ ($fm^{-2}$)  &  $(g_\delta/m_\delta)^2$ ($fm^{-2}$) &  $S_0$ (MeV)  & L (MeV)   \\
 \hline
 14.687 & 3.00  & 32.49 & 118.4   \\
 \hline
 12.991 & 2.50 & 32.49 & 113.9 \\
 \hline
 11.287 & 2.00 & 32.49 & 109.6  \\
\hline
 9.578 & 1.50 & 32.49 & 105.4  \\
\hline  
 7.858 & 1.00 & 32.49 & 98.5  \\
\hline
 4.410 & 0.00 & 32.49 & 93.9  \\
\hline
\end{tabular} 
\caption{ $L$ values  with GM1 for a fixed symmetry energy at 32.49 MeV} 
\label{T10}
\end{center}
\end{table}

\begin{table}[!htb]
\begin{center}
\begin{tabular}{|c|c||c|c|}
\hline 
$(g_\rho/m_\rho)^2$ ($fm^{-2}$)  &  $(g_\delta/m_\delta)^2$ ($fm^{-2}$) &  $S_0$ (MeV)  & L (MeV)   \\
 \hline
 13.620 & 2.50  & 32.49 & 103.3   \\
 \hline
 11.864 & 2.00 & 32.49 & 100.4 \\
 \hline
 10.103 & 1.50 & 32.49 & 97.5  \\
\hline
 8.336 & 1.00 & 32.49 & 94.8  \\
\hline  
 4.791 & 0.00 & 32.49 & 89.7  \\
\hline
\end{tabular} 
\caption{$L$ values  with GM3 for a fixed symmetry energy at 32.49 MeV} 
\label{T11}
\end{center}
\end{table}

\begin{table}[!htb]
\begin{center}
\begin{tabular}{|c|c||c|c|}
\hline 
$(g_\rho/m_\rho)^2$ ($fm^{-2}$)  &  $(g_\delta/m_\delta)^2$ ($fm^{-2}$) &  $S_0$ (MeV)  & L (MeV)   \\
 \hline
 12.444 & 2.50  & 33.40 & 139.2   \\
 \hline
 10.828 & 2.00 & 33.40 & 132.1 \\
 \hline
 9.204 & 1.50 & 33.40 & 125.3  \\
\hline
 7.571 & 1.00 & 33.40 & 118.8  \\
\hline  
 4.280 & 0.00 & 33.40 & 106.9  \\
\hline
\end{tabular} 
\caption{$L$ values  with NL3 for a fixed symmetry energy  at 33.40 MeV} 
\label{T12}
\end{center}
\end{table}

\begin{table}[!htb]
\begin{center}
\begin{tabular}{|c|c||c|c|}
\hline 
$(g_\rho/m_\rho)^2$ ($fm^{-2}$)  &  $(g_\delta/m_\delta)^2$ ($fm^{-2}$) &  $S_0$ (MeV)  & L (MeV)   \\
 \hline
 12.795 & 2.50  & 31.49 & 105.1   \\
 \hline
 11.064 & 2.00 & 31.49 & 101.4 \\
 \hline
 9.331 & 1.50 & 31.49 & 97.9  \\
\hline
 7.589 & 1.00 & 31.49 & 94.5  \\
\hline  
 4.070 & 0.00 & 31.49 & 88.0  \\
\hline
\end{tabular} 
\caption{$L$ values  with NL$\rho$ for a fixed symmetry energy at 31.49 MeV} 
\label{T13}
\end{center}
\end{table}

If the symmetry energy is fixed, the $\delta$ meson  always  forces the
slope to increase, as pointed in Ref.~\cite{Liu}. We also see that, although we are 
able to always construct an EoS with reasonable values of symmetry energy,
some parametrizations have  too large values of $L$ ($L$ $>$ 113 MeV). We keep these
parametrizations for the sake of comparison.


\begin{figure*}[hb]
\begin{tabular}{cc}
\includegraphics[width=5.6cm,height=6.2cm,angle=270]{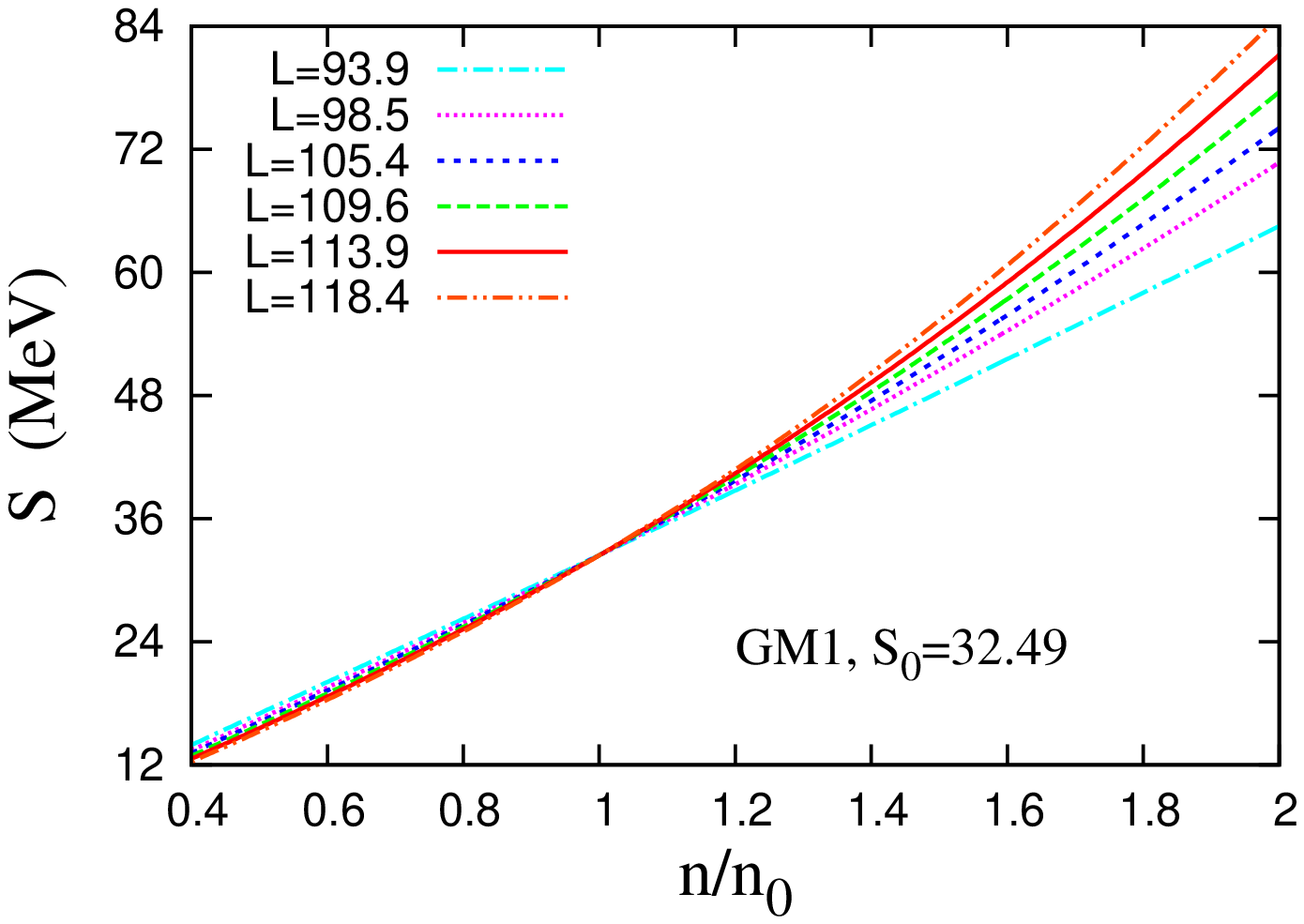} &
\includegraphics[width=5.6cm,height=6.2cm,angle=270]{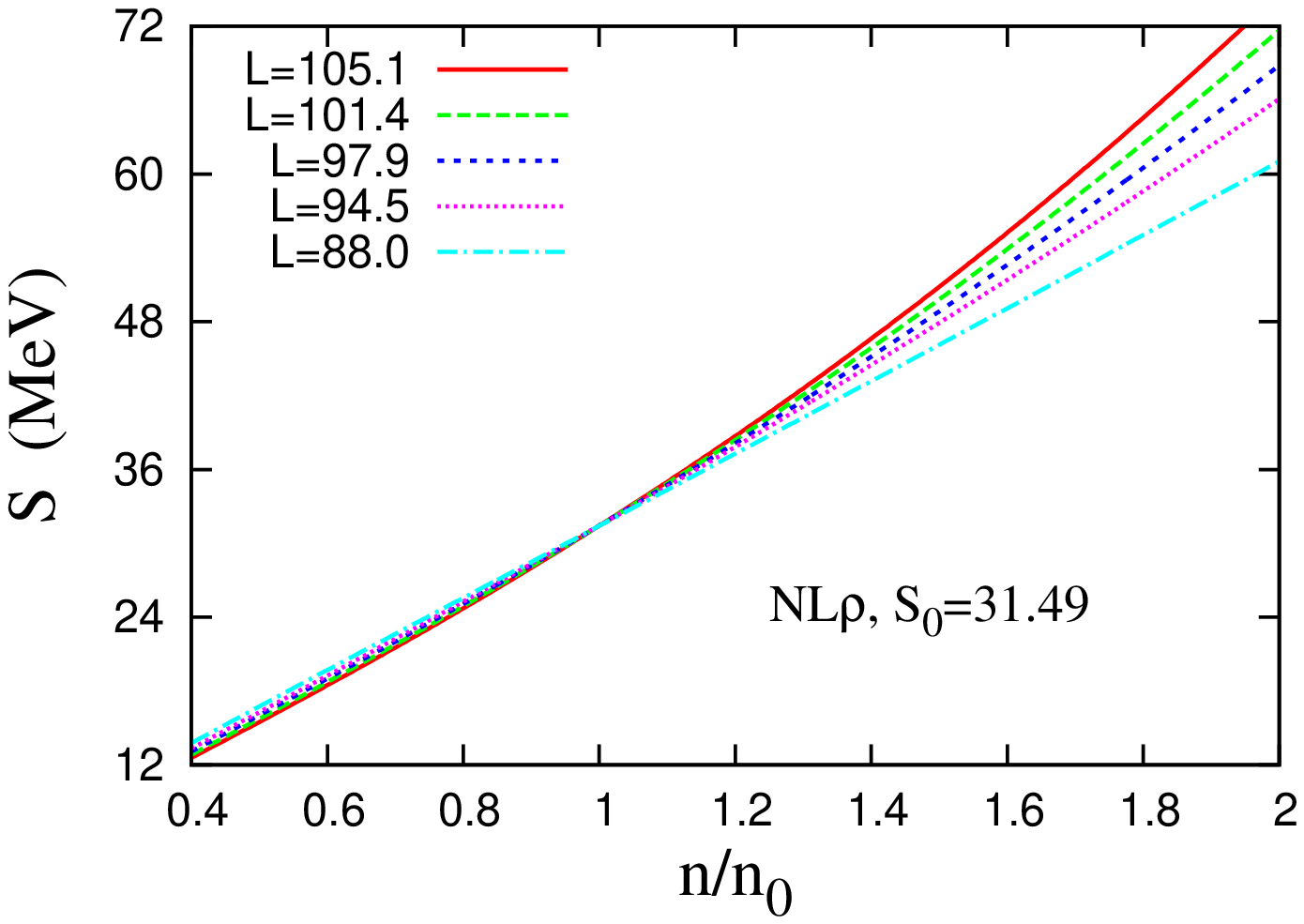} \\
\includegraphics[width=5.6cm,height=6.2cm,angle=270]{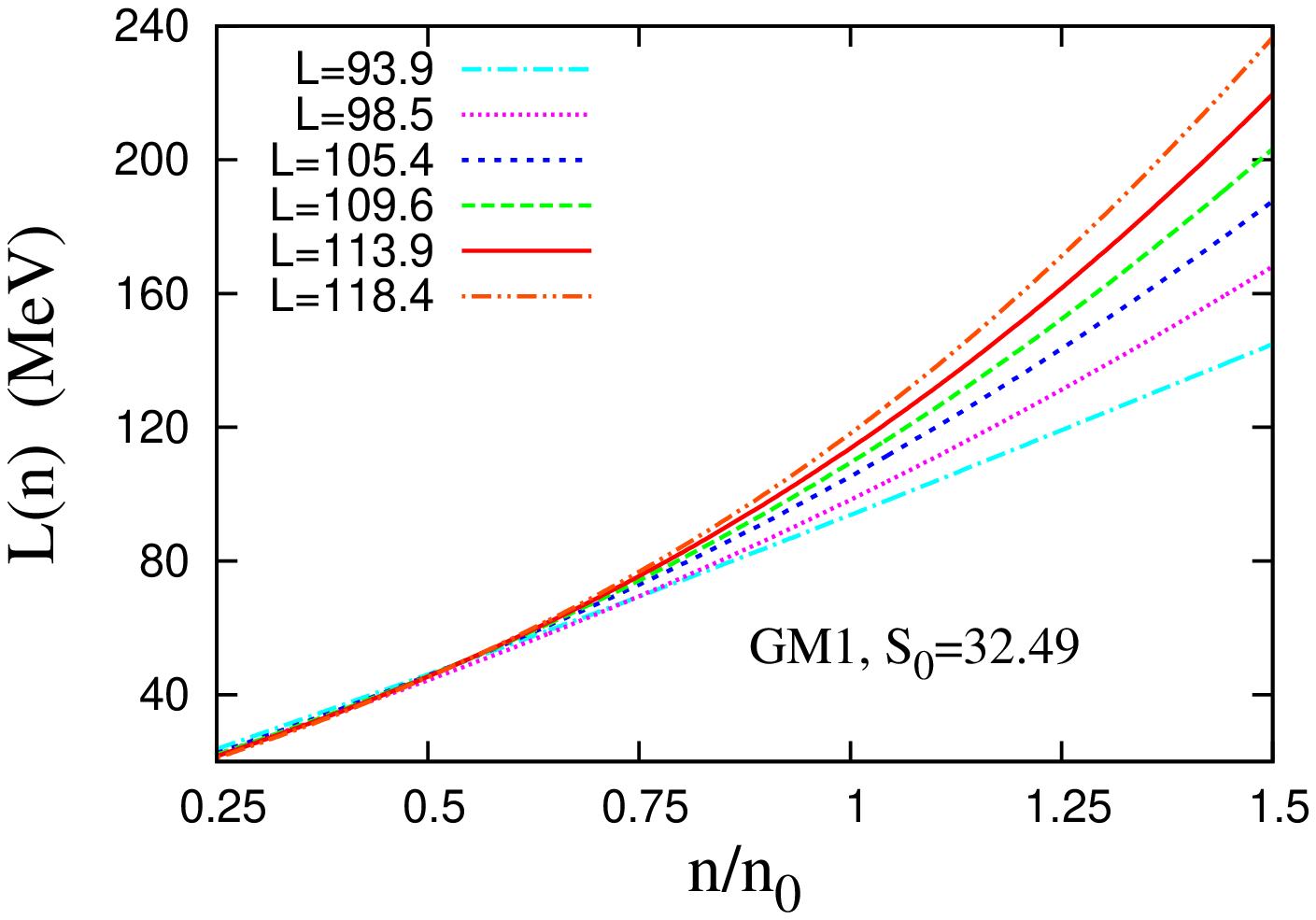} &
\includegraphics[width=5.6cm,height=6.2cm,angle=270]{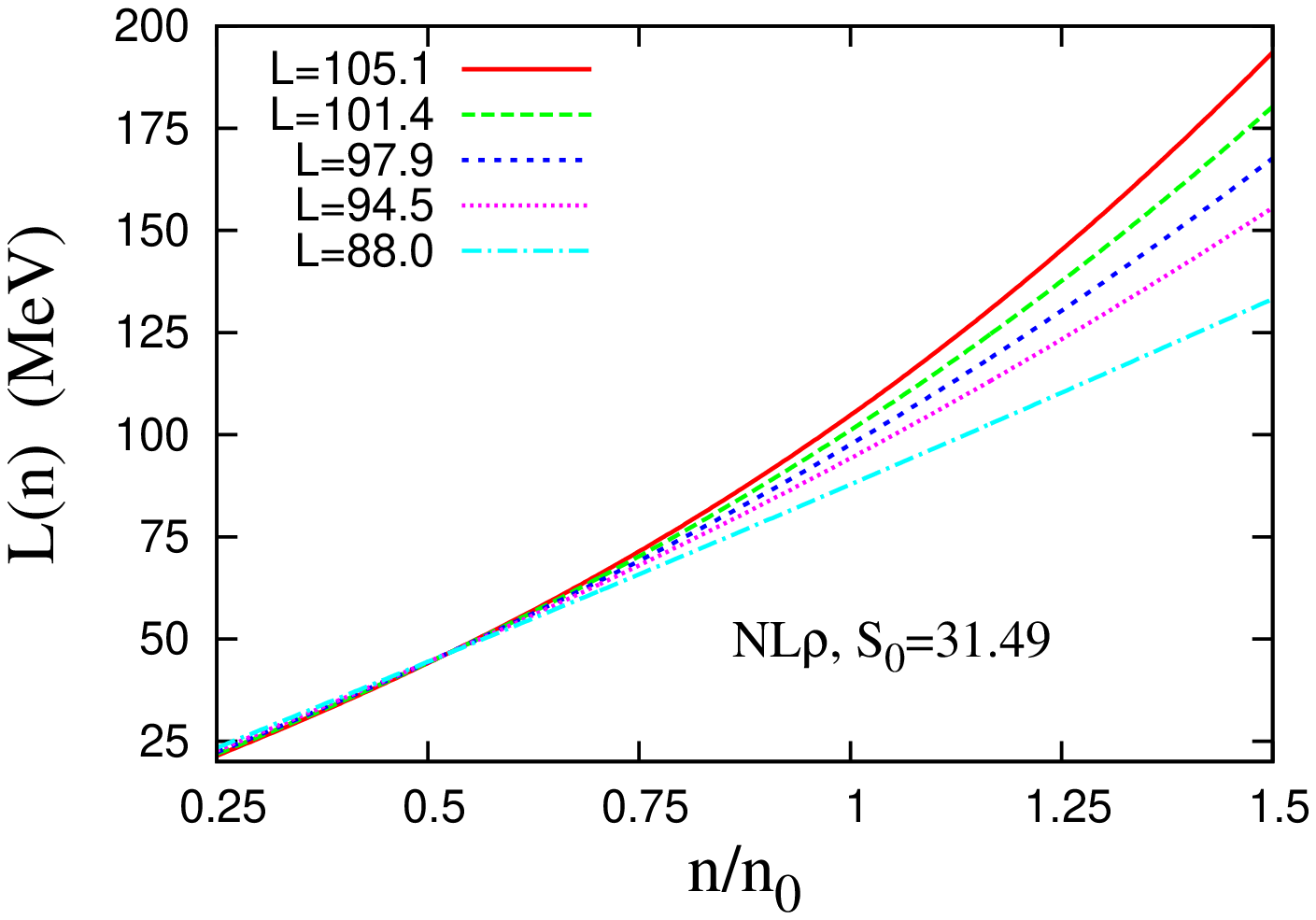} \\
\end{tabular}
\caption{(Color online) (Top) Symmetry energy $S$ as function of density and (Bottom) slope of the
symmetry energy $L(n)$ as function of density with fixed $S_0$. } \label{F11}
\end{figure*}

We plot the density dependent symmetry energy $S$ and the
slope $L(n)$ for this approach in Fig. \ref{F11} for GM1 and
NL$\rho$ parametrization. 
Fixing $S_0$ rather than  fixing $L$ causes the crossing of the curves to take place at lower densities.
When we fix the slope $L$, the crossing of $L(n)$, which obviously always happened at $n = n_0$ in the 
previous case,
now happens at densities around $0.6n_0$, while the symmetry energy crossing, which happened
around $1.8n_0$, now obviously happens at $n=n_0$.

 \begin{figure*}[hb]
\begin{tabular}{cc}
\includegraphics[width=5.6cm,height=6.2cm,angle=270]{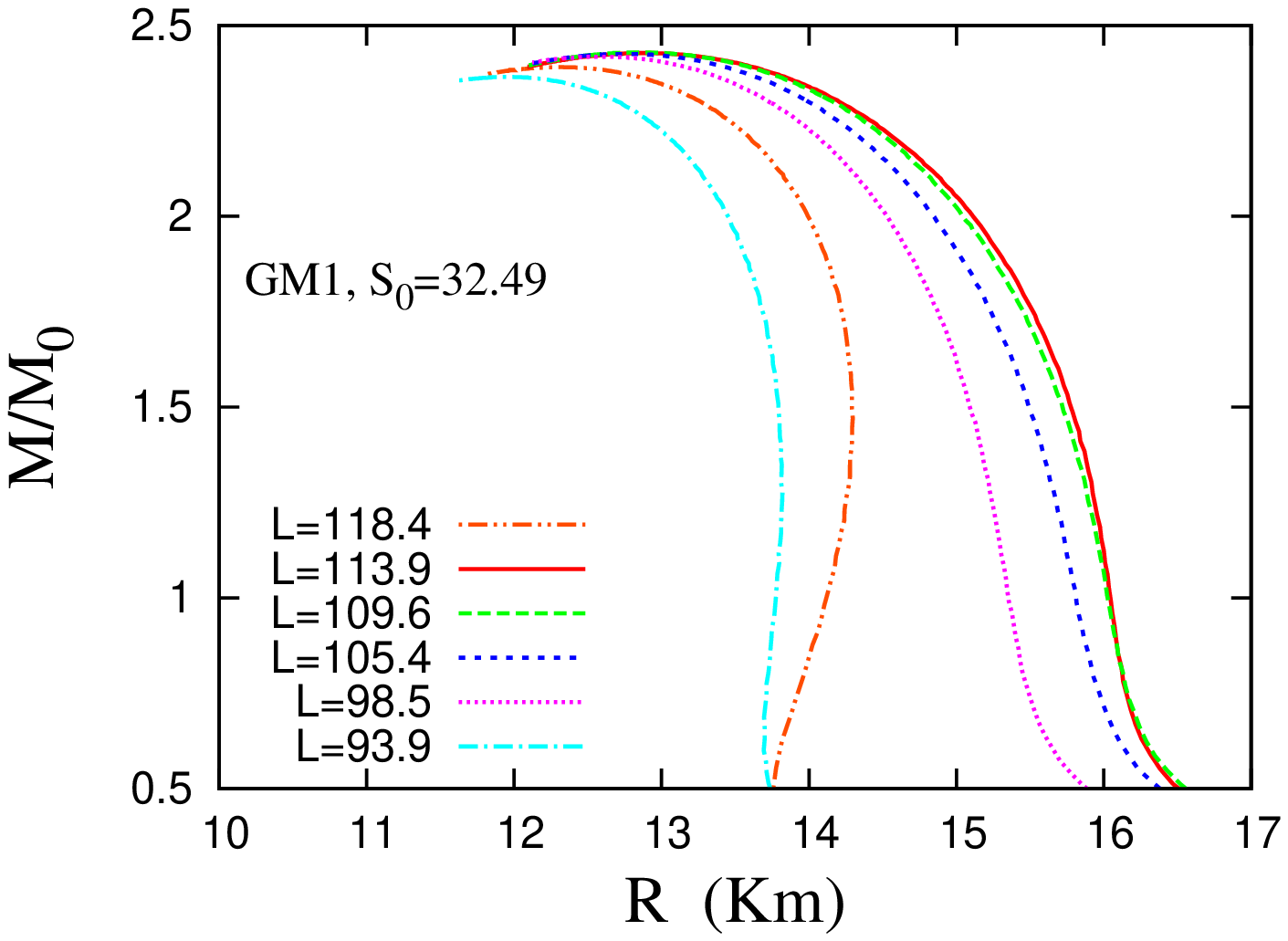} &
\includegraphics[width=5.6cm,height=6.2cm,angle=270]{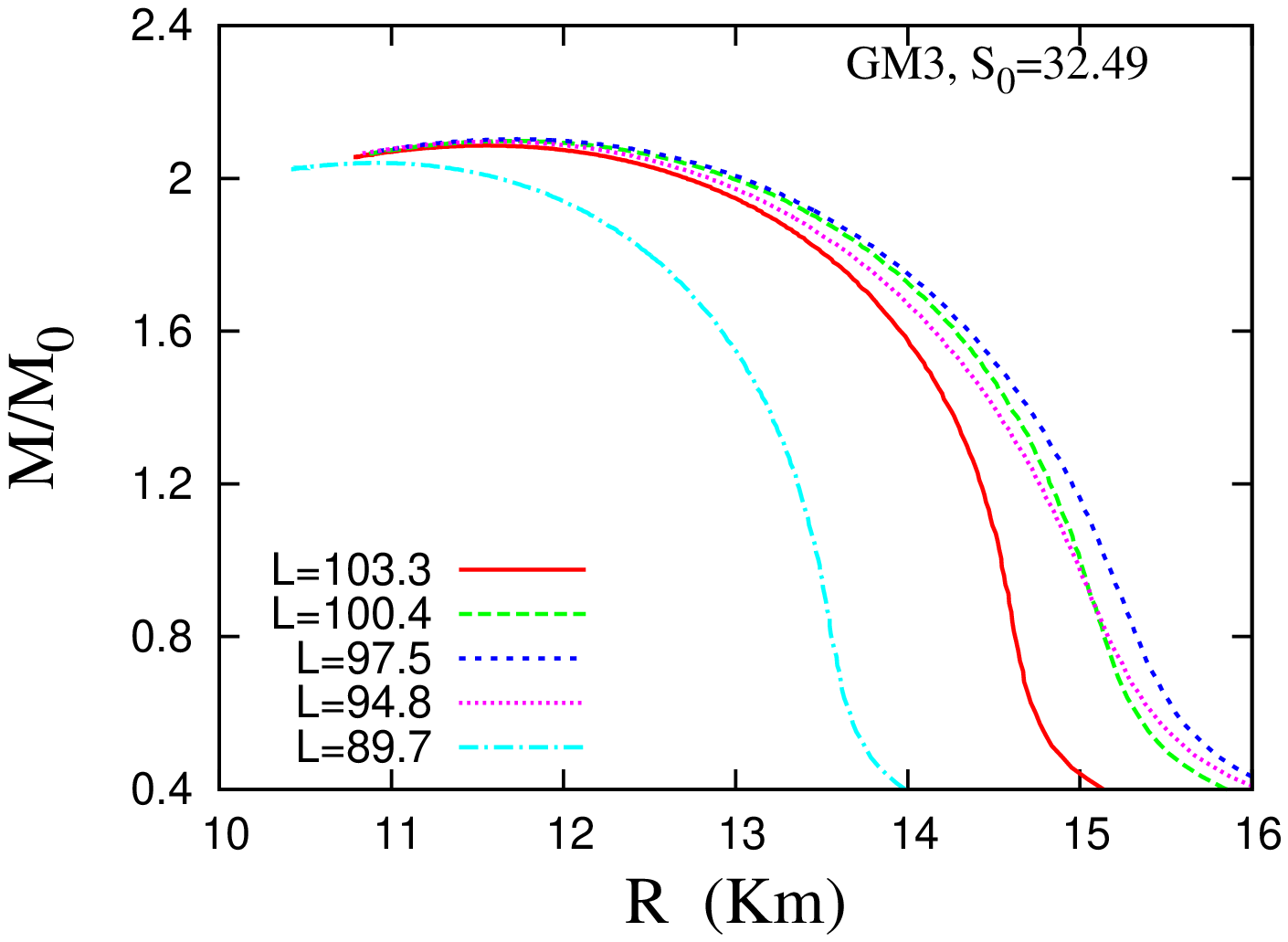} \\
\includegraphics[width=5.6cm,height=6.2cm,angle=270]{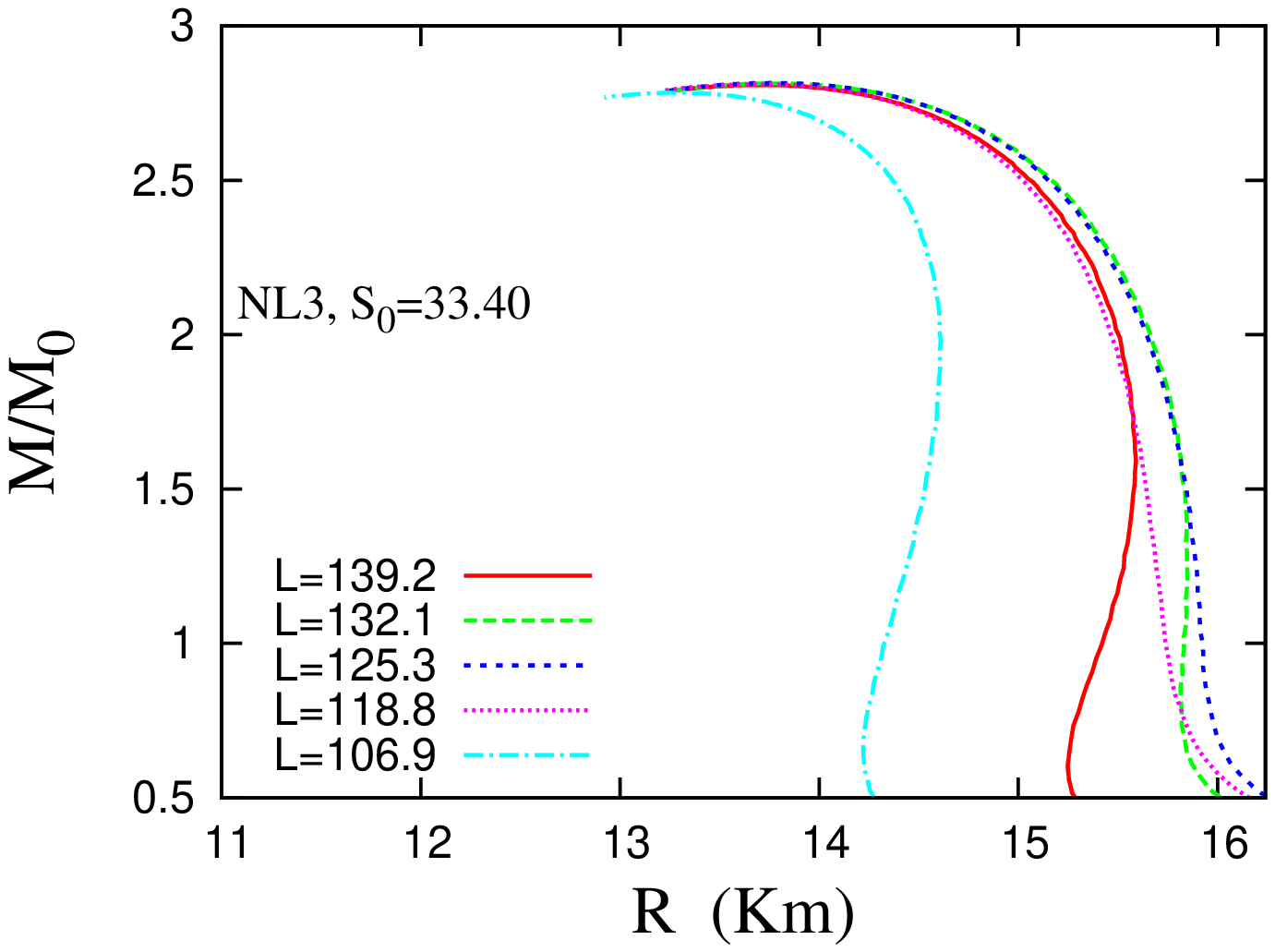} &
\includegraphics[width=5.6cm,height=6.2cm,angle=270]{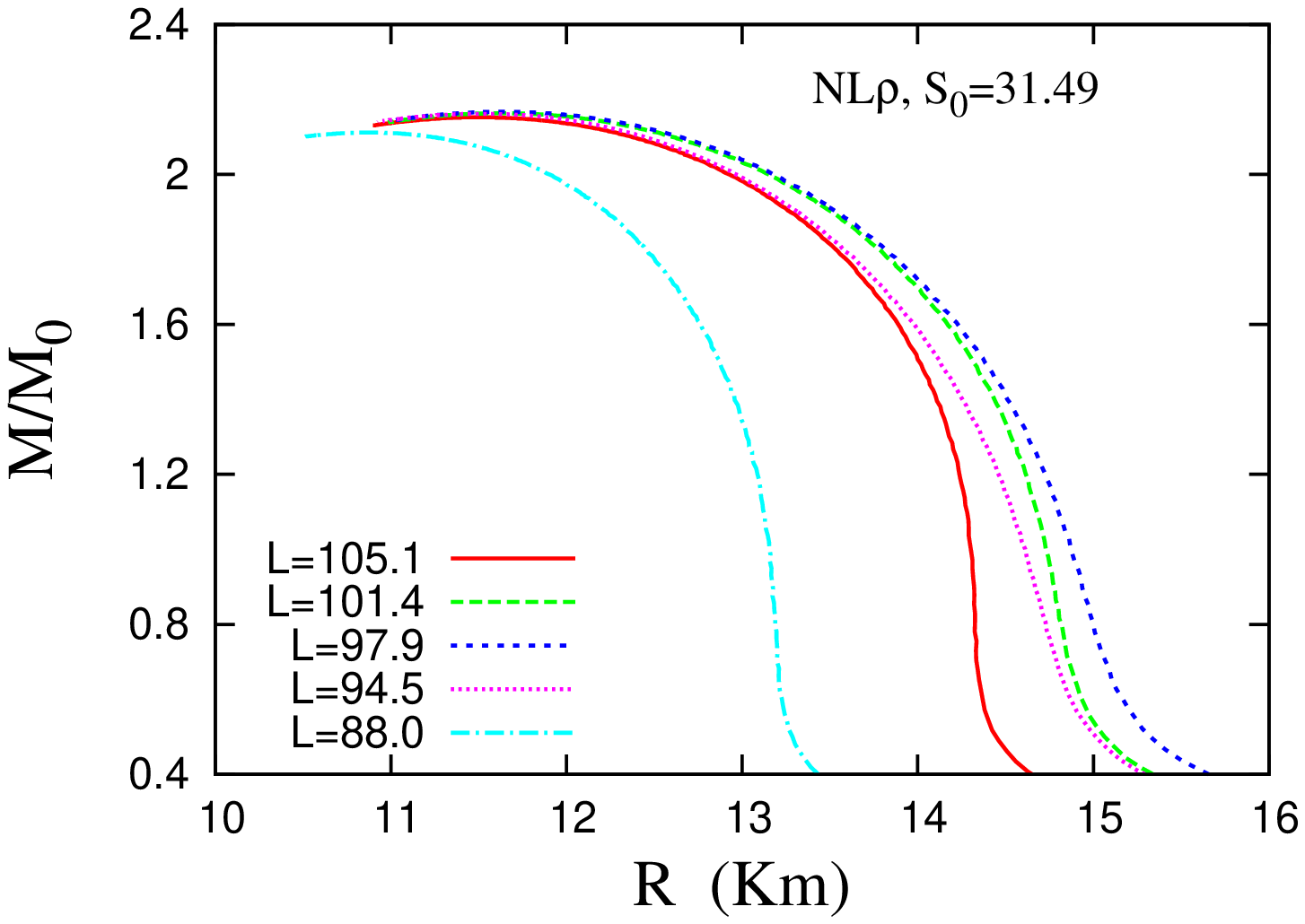} \\
\end{tabular}
\caption{(Color online) Neutron star mass-radius relation with fixed $S_0$. } \label{F13}
\end{figure*}

We plot the TOV solution in Fig. 13.
The effect of the slope $L$ on neutron star properties was already studied in previous 
works~\cite{Rafa2011,Debdelta,Liu2}.
Increasing the slope, the radii of canonical 1.4$M_\odot$ increases, as pointed out in ref.~\cite{Rafa2011}. 
However, we can see
that there is a limit. If the slope increases too much, the radii begins to drop again. We found that 
there is a
maximum value of the neutron star radius. This maximum possible value could also indicate
that there is a theoretical superior limit of acceptable  values for the slope $L$ 
for a fixed symmetry energy $S_0$.
Moreover, as pointed out earlier, the fact that a fixed $L$ produces differences of up to 1.7 km
on the radii of the canonical 1.4$M_\odot$ indicates that although the slope gives us insight about the radii
it is not enough to fully determine it. And again, the symmetry energy and its slope have almost no
influence on the maximum mass of the neutron star.

 \begin{figure*}[hb]
\begin{tabular}{cc}
\includegraphics[width=5.6cm,height=6.2cm,angle=270]{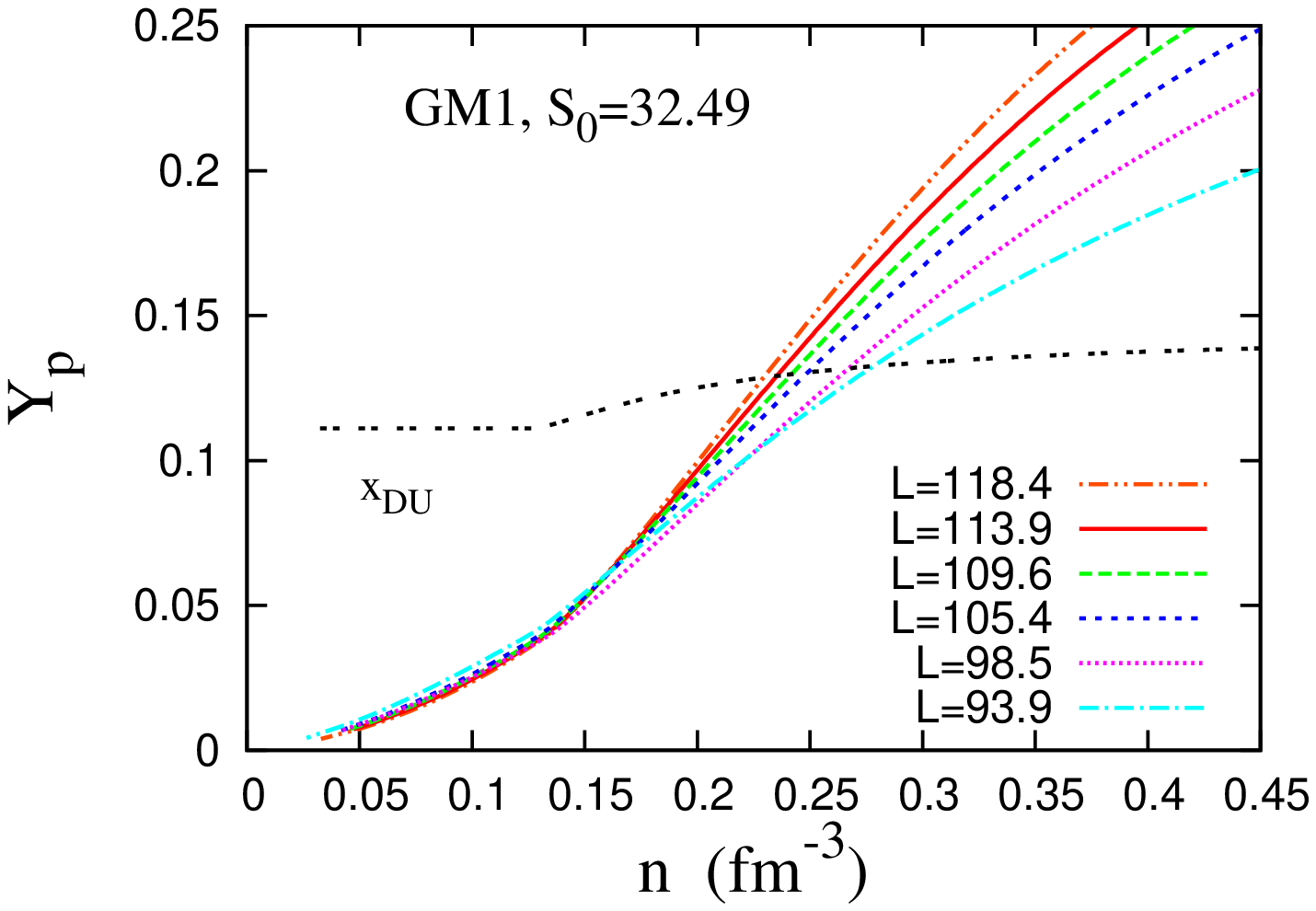} &
\includegraphics[width=5.6cm,height=6.2cm,angle=270]{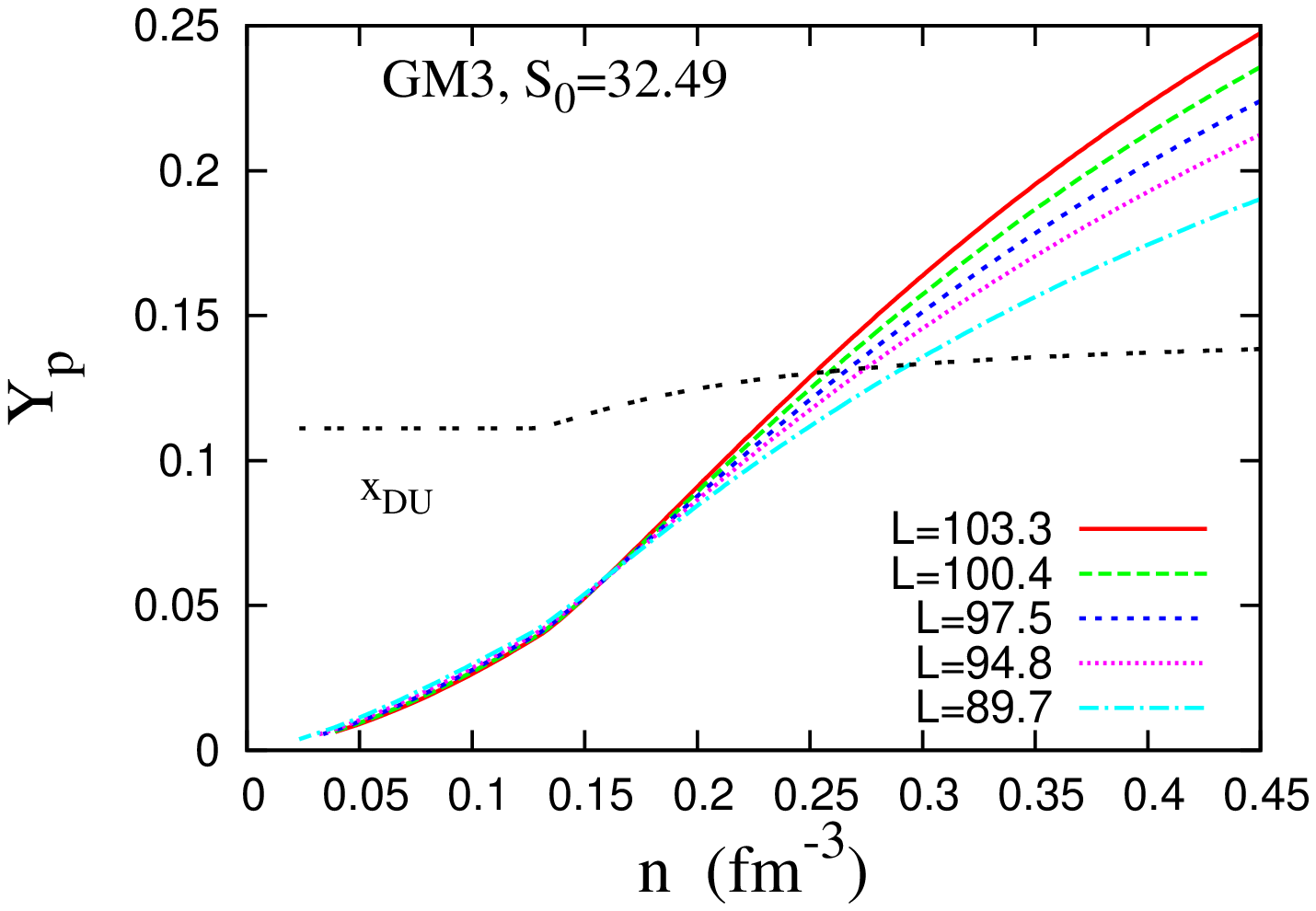} \\
\includegraphics[width=5.6cm,height=6.2cm,angle=270]{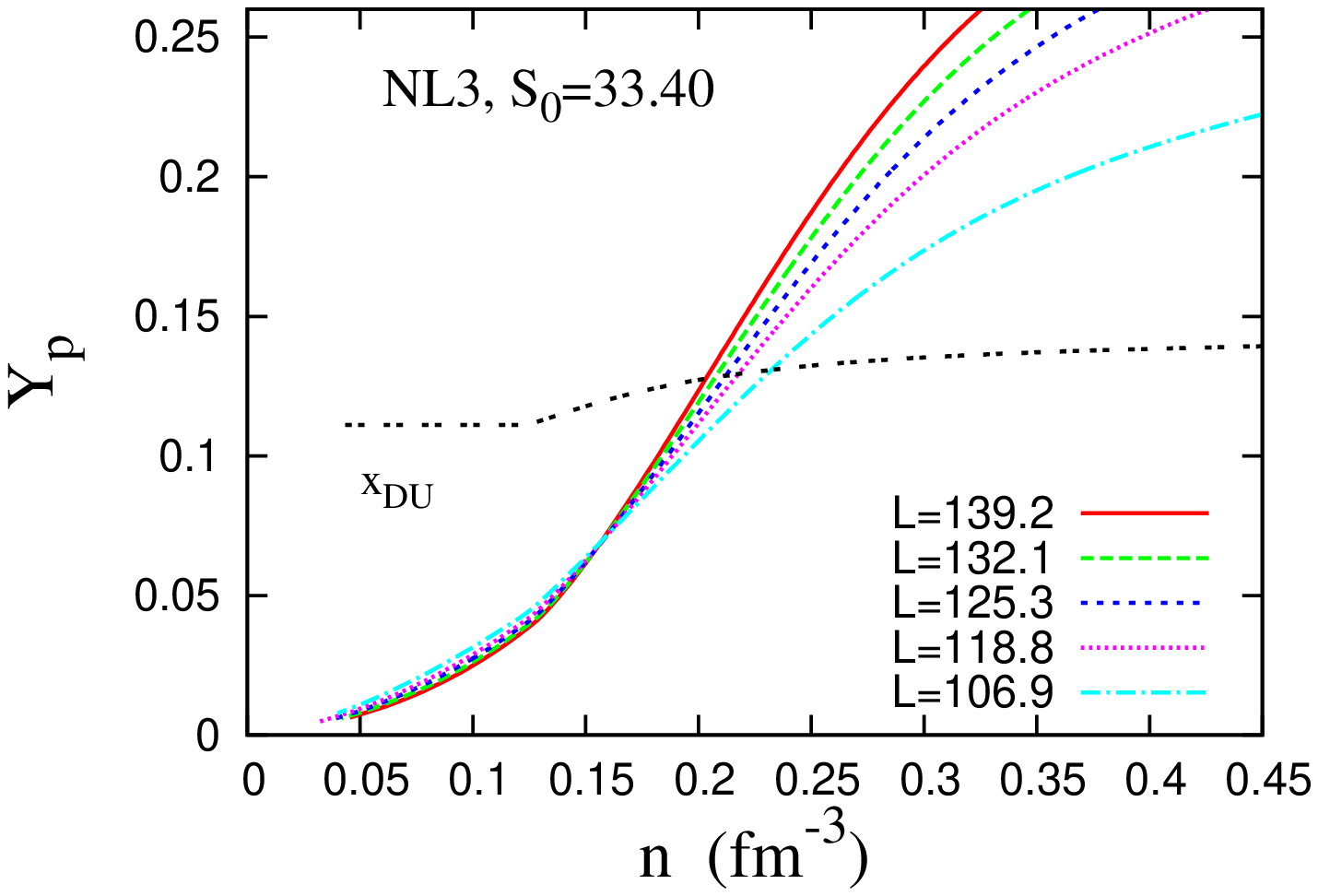} &
\includegraphics[width=5.6cm,height=6.2cm,angle=270]{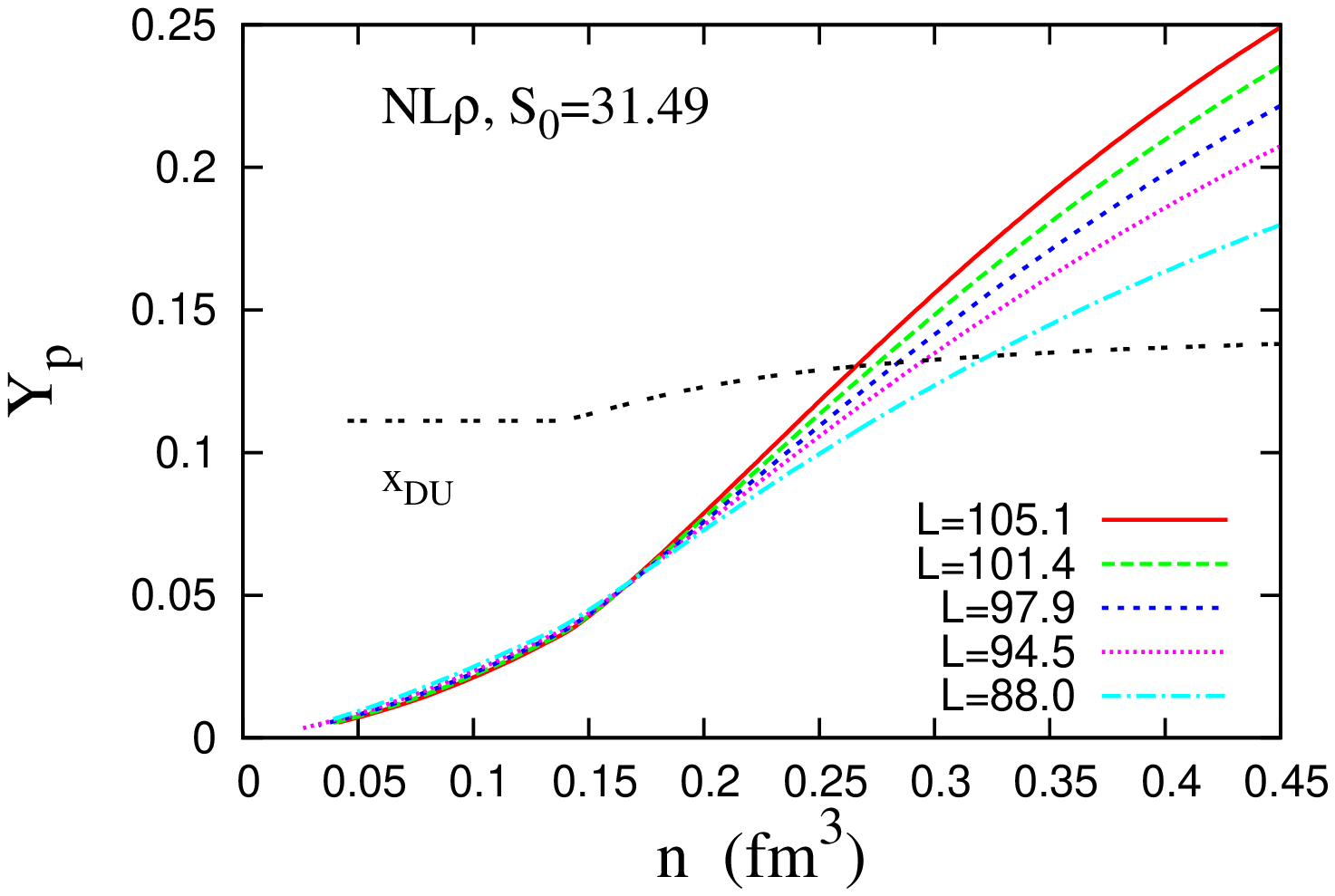} \\
\end{tabular}
\caption{(Color online) Proton fraction $Y_p$ and  the critical  value $x_{DU}$
 with fixed $S_0$.  } \label{F14}
\end{figure*}

Now we return to the DU process and plot the proton fraction in Fig. \ref{F14}.
Fixing the symmetry energy $S_0$ causes the crossing to happens for densities
below those which enable DU process.
This is expected since this behaviour was already found in
 the density dependent symmetry energy  $S_0$, and its slope $L(n)$.

\begin{figure}[hb] 
\begin{centering}
 \includegraphics[angle=270,
width=0.4\textwidth]{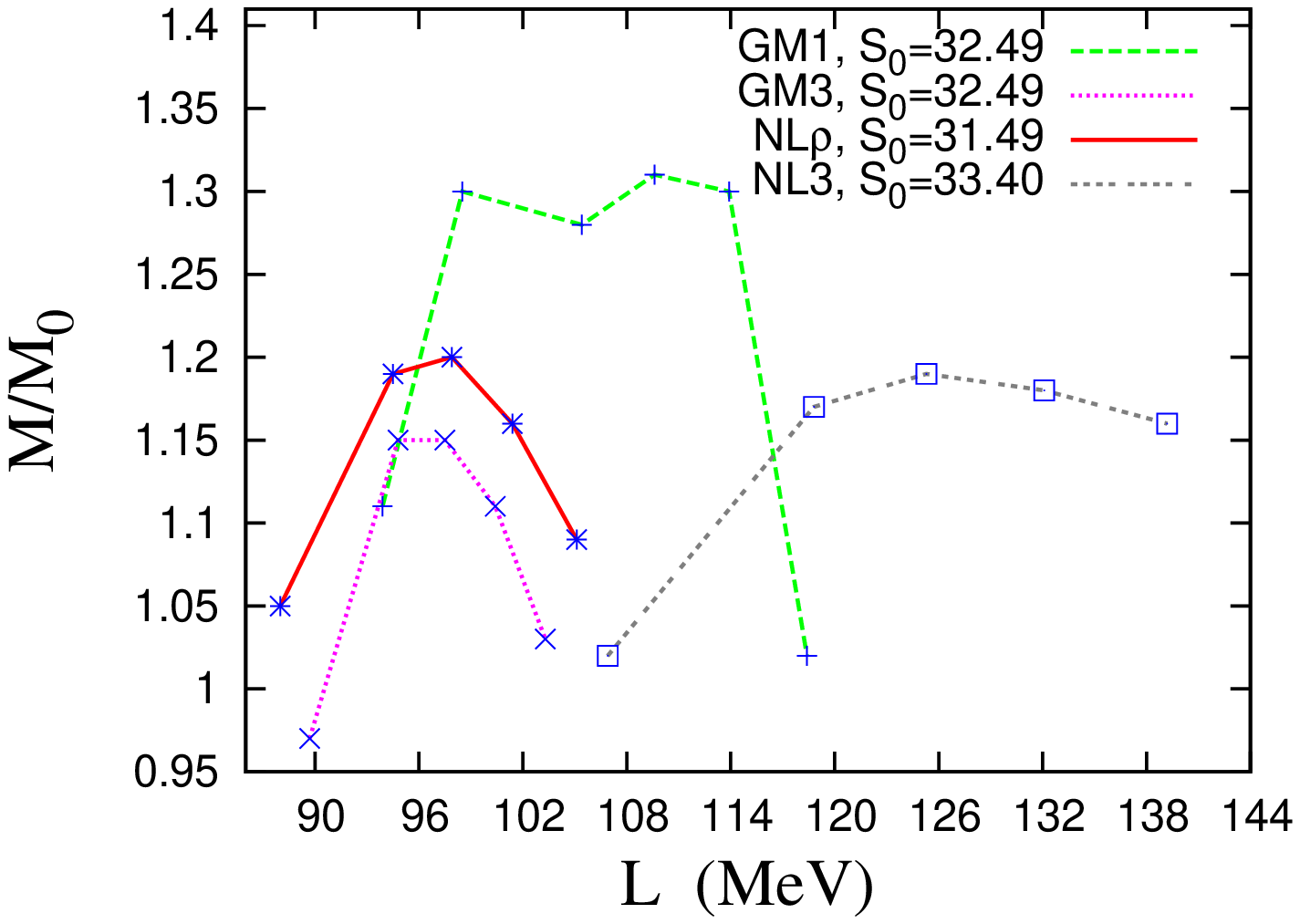}
\caption{(Color online) Minimum mass that enable DU process with fixed $S_0$.} \label{F15}
\end{centering}
\end{figure}

The minimum mass that enables DU process for fixed symmetry energy
is plotted in Fig. \ref{F15} from where we see that
there is  an oscillation in the mass, that
could again  be associated with a too large value of $L$.
It is interesting to note that although the NL3 has a higher value of maximum mass,
GM1 produces a larger minimum mass  which enables DU process.

The main results are shown in Table \ref{T16}.

\begin{table}[!htb]
\begin{center}
\begin{tabular}{|c|c|c|c|c|c|}
\hline 
 Model &  $L$ (MeV) & $M_{max}/M_\odot$  & $R_{1.4M_\odot}$ & $M_{DU}/M_\odot$ & $n_{DU}$ ($fm^{-3}$)   \\
 \hline
 GM1 & 93.9  & 2.39 & 13.80 & 1.10 & 0.279   \\
 \hline
GM1 &  98.5 & 2.43 & 15.17 & 1.30 &  0.267 \\
 \hline
 GM1 & 105.4 & 2.44 & 15.58 & 1.28 & 0.250 \\
\hline
 GM1 & 109.6 & 2.44 & 15.81 & 1.31 & 0.243  \\
\hline  
 GM1 & 113.9 & 2.44 & 15.85 & 1.30 & 0.237  \\
\hline  
 GM1 & 118.4 & 2.40 & 14.29 & 1.02 & 0.232  \\
\hline
\hline
GM3 & 89.7 & 2.04 & 13.16 & 0.98 & 0.293 \\
\hline
GM3 & 94.8 & 2.09 & 14.48 & 1.15 & 0.275 \\
\hline
GM3 & 97.5 & 2.10 & 14.68 & 1.15 & 0.267 \\
\hline
GM3 & 100.4 & 2.09 & 14.58 & 1.11 & 0.260 \\
\hline
GM3 & 103.3 & 2.07 & 14.23 & 1.03 & 0.254 \\
\hline
\hline
NL3 & 106.9 & 2.80 & 14.54 & 1.04 & 0.232 \\
\hline
NL3 & 118.8 & 2.83 & 15.65 & 1.17 & 0.219 \\
\hline
NL3 & 125.3 & 2.83 & 15.86 & 1.19 & 0.213  \\
\hline
NL3 & 132.1 & 2.83 & 15.84 & 1.18 & 0.209 \\
\hline
NL3 & 139.2 & 2.82 & 15.56 & 1.16 & 0.206 \\
\hline
\hline
NL$\rho$  & 88.0 & 2.11 & 12.97 & 1.05 & 0.323 \\
\hline
NL$\rho$ & 94.5 & 2.16 & 14.24 & 1.19 & 0.298 \\
\hline
NL$\rho$ & 97.9 & 2.17 & 14.49 & 1.20 & 0.285 \\
\hline
NL$\rho$  & 101.4 & 2.16 & 14.43 & 1.16 & 0.276 \\
\hline
NL$\rho$ & 105.1 & 2.15 & 14.10 & 1.09 & 0.269 \\
\hline
\end{tabular} 
\caption{Neutron star main properties with fixed  $S_0$.} 
\label{T16}
\end{center}
\end{table}

Finally we return to the question of the radii of the canonical 1.4$M_\odot$.
We pointed out that the slope alone does not give us enough information
to determine them with precision. We rise a hypothesis that it is not the
slope $L$ but the variation of $L(n)$ at sub-threshold densities  that is correlated
with the neutron star radii.

\begin{figure}[hb] 
\begin{centering}
 \includegraphics[angle=270,
width=0.4\textwidth]{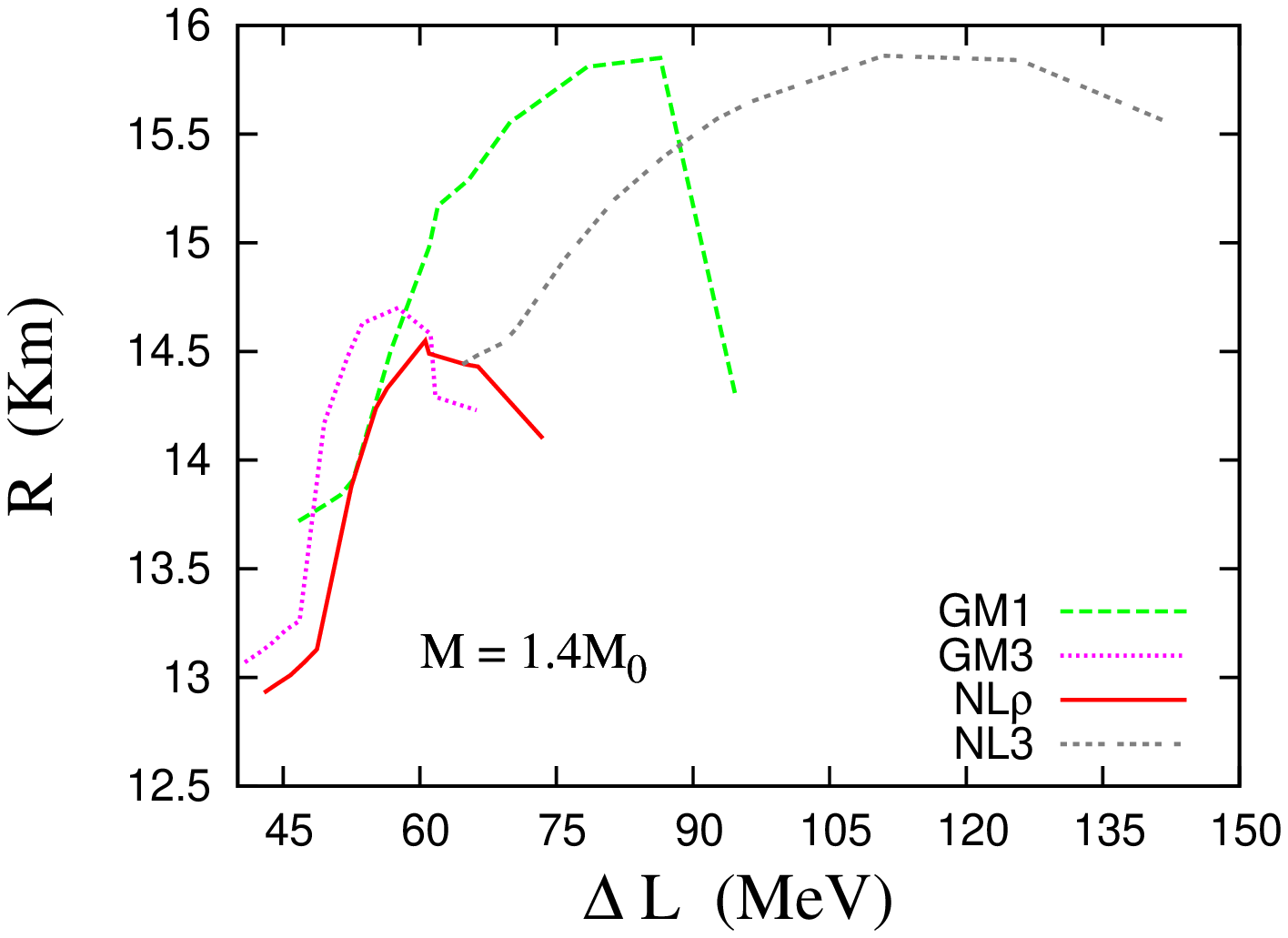}
\caption{(Color online) Neutron star radii as function of the variation of the slope $\Delta L$. } \label{F16}
\end{centering}
\end{figure}

We have arbitrarily calculated the value $ \Delta L = L(1.25) - L(0.75)$ and plotted
the results of all three approaches in Fig. \ref{F16}.

We see that the neutron stars radii grow with  $\Delta L$, reach a maximum, and drop again.
This explains the fact that the same slope causes variations of up to 1.7 km in the radii and 
the oscillation in the radii with the increase of the slope. We can also see that there is a maximum value of
the radii that depends on $\Delta L$.

Moreover, this could indicate that the values of $\Delta L$ beyond the maximum radius are too large,
limiting the maximum value for the slope $L$. For GM1 and NL3 these values are 
113.9 MeV and 125.3 MeV for $S_0$ equal to 32.49 MeV and 33.40 MeV respectively.
The NL3 value is ruled out, since experiments point to a maximum value
of 113 MeV~\cite{Tsang,Chen} { and not larger than 115 MeV \cite{Dutra} and the GM1 value is too 
close to the accepted limit.}
However, for GM3 and NL$\rho$ the  values of $L$ 
beyond the maximum radii are 97.5 MeV and 97.9 MeV for $S_0$ equal to 32.49 MeV and
31.49 MeV respectively.  The results can be even lower.  For $S_0$ equal
to 30.15 MeV and 30.04 MeV in GM3 and NL$\rho$ respectively, the values of $L$, 
which respectively are 95.7 MeV and 97.0 MeV, yielding values of $\Delta L$ that are
 already beyond the maximum radii, indicating
a relatively low  theoretical superior limit for the slope $L$.


\section{Final remarks and conclusions}

In this work we discuss the role of the symmetry energy and the slope 
on neutron star properties. To accomplish that, we divide our study in 
three different fronts, varying both $S_0$ and the related $L$ in the traditional
$\sigma\omega\rho$ model, fixing $L$ and varying $S_0$ within the $\sigma\omega\rho\delta$
model, and vice-versa. We then obtain  the maximum masses of neutron star,
the radii of the canonical 1.4 $M_\odot$, and the minimum mass that enables direct
URCA process for each case.

We see that the maximum mass is not very much influenced by both $S_0$ and $L$
and the differences are never larger than $0.06M_\odot$. 
 Since the EoS of $\beta$ stable matter is very little sensitive to the variations 
of the symmetry energy and its slope~\cite{Rafa2011,Debdelta},
 we do not plot these graphs, which can be easily
 found in the literature~\cite{Glen,Debdelta,Liu2,Paoli,Luiz}.

We have confirmed that the radii of the canonical 1.4 $M_\odot$ is not
affected by the symmetry energy $S_0$, whereas in some cases the radii  increase with it, while in others, 
there is a decrease. Nevertheless, the radii is correlated with  a  variation of the slope $\Delta L$.
 The radii increase with $\Delta L$ up to a maximum value, then 
drop again. This behaviour can be associated with a maximum theoretical value of $L$, and 
provide a { possible} constraint to nuclear matter. Our models predict radii from 12.9 km to 15.9 km. We are unable to explain  the
large pulsars as RX J1856.5-3754~\cite{Wynn} with  radius of 17 km, neither the very small ones as
those pointed out in Ref.~\cite{Steiner,guillot} with radii lying between 9-12 km.
Neutron stars with small radii, are generally related
to  very low slope $L$, as suggested in some works~\cite{Steiner,Lim}.
There is a QHD model that reproduces a low slope, the FSU~\cite{Pie1}.
However, previous studies~\cite{Rafa2011,Shen} indicate that the maximum mass within the FSU model is  only 1.7 $M_\odot$,
and in the light of the recent super massive pulsars~\cite{Demo,Antoniadis}, the FSU model cannot represent
the ultimate EoS of  nuclear matter. These controversial results alongside the possibility of a slope as high as
170 MeV~\cite{Cozma} led us to  believe that the neutron star radii is still an open puzzle.

Finally we study the influence of the symmetry energy and its slope on direct URCA process, which is directly related
to the proton fraction $Y_p$. The minimum mass that enables DU process  changes dramatically
with the symmetry energy $S_0$. Indeed the variation of the minimum mass can reach more than $50\%$.
Studies of neutron star cooling~\cite{Pieka,Yakov} should be used  to
constrain  nuclear matter.

We next discuss the validity of our models in the light of experimental
constraints of nuclear matter besides the symmetry energy and the slope,
based in the discussion presented in Ref.~\cite{Dutra}. 
The first physical quantity checked in~\cite{Dutra} is the compressibility
$K$, where the experimental results point to  values between 190 - 270 MeV.
This implies that while GM3 and $NL\rho$ fulfil this constraint, 
 GM1 and NL3 have to be used with care, since both have
values of $K$ higher than 270 as pointed out in Table \ref{T1}.
{ Recently, new constraints from re-analysis of data on GMR energies 
\cite{youngblood2004, uchida} became available, suggesting that the compressibility values
could move to somewhat higher values, 250 - 315 MeV. If this is 
confirmed, then GM3 and NL$\rho$ are the parametrizations that would be slightly
out of the desired range.}

The second physical quantity is the pressure of symmetric matter
up to densities  five times the  nuclear saturation density, based
on Ref.~\cite{Daniel}. This constraint was already studied in 
other works~\cite{Rafa2011,Dutra,Luiz2}, and again,
 while it is satisfied by the GM3 and NL$\rho$, 
GM1 and NL3 are outside the experimental region
of the pressure. However, as pointed in Ref.~\cite{Luiz2}, 
GM1 can fulfil this constraint if we assume that hyperons
populates the nuclear matter at high density.

In Ref.~\cite{Dutra},  the  $S_0$ and $L$ values are also discussed.
$S_0$ is assumed to lie between 30 - 35 MeV, exactly as indicated
in ref.~\cite{Tsang,Sun,Carbone}; and the maximum value of $L$  being
 115 MeV, which is very close of 113 MeV~\cite{Tsang,Chen}
used in our work.  

Nuclear astrophysics also provide constraints for nuclear matter.
For instance, the discovery of two supermassive pulsars~\cite{Demo,Antoniadis}
indicates a moderate stiff EoS for high density. Moreover, the possible
existence of a 2.7 solar mass, correspondent to the 
pulsar PSR J1311-3430~\cite{Romani}, indicates a very stiff EoS.

Since the $\delta$ scalar-isovector meson always increase the slope~\cite{Liu}, in this work we have studied 
models with high values of $L$. The next step of our work is study models with low slope adding a
non linear $\omega-\rho$ coupling as in the FSU parametrization ~\cite{Pie1}. 
Although the FSU model seems to fail the description of
super massive pulsars, we can add this term in the standard parametrizations, as we did with the $\delta$
meson, and check if there is analogous results as a minimum neutron star radii.

Moreover, at high densities, strange content particles
can be created~\cite{Glen}. However, they bring  several ambiguities since we
do not know the strength of the hyperon-meson interaction~\cite{Luiz2,Weiss2}
implying that the maximum mass can vary up to 100$ \% $~\cite{Glen}. Also, when hyperons
are present,  strange  mesons could be important  to mediate the
hyperon-hyperon interactions~\cite{Luiz2,Weiss2,Elis,Rafa2005}. 
Alongside hyperons, strong magnetic fields~\cite{Peng}, 
strongly affects the macroscopic properties of the neutron star. 
 Works along these lines are in progress.

\vspace{.5cm}

$\mathbf{Acknowledgments}$

\vspace{.5cm}
This work was partially supported by CNPq (Brazil), CAPES (Brazil) and FAPESC (Brazil) under project 
2716/2012,TR 2012000344.

\end{document}